# Examining the development of attitude scales using Large Language Models (LLMs)

**Maria Symeonaki[*], Giorgos Stamou[**], Aggeliki Kazani[*], Eva Tsouparopoulou[*], Glykeria Stamatopoulou[*]**

[*]School of Political Sciences, Department of Social Policy
Panteion University of Social and Political Sciences

[**]School of Electrical and Computer Engineering
National Technical University of Athens

**Abstract**

For nearly a century, social researchers and psychologists have debated the efficacy of psychometric scales for attitude measurement, focusing on Thurstone's equal appearing interval scales and Likert's summated rating scales. Thurstone scales fell out of favour due to the labour-intensive process of gathering judges' opinions on the initial items. However, advancements in technology have mitigated these challenges, nullifying the simplicity advantage of Likert scales, which have their own methodological issues. This study explores a methodological experiment to develop a Thurstone scale for assessing attitudes towards individuals living with AIDS. An electronic questionnaire was distributed to a group of judges, including undergraduate, postgraduate, and PhD students from disciplines such as social policy, law, medicine, and computer engineering, alongside established social researchers, and their responses were statistically analysed. The primary innovation of this study is the incorporation of an Artificial Intelligence (AI) Large Language Model (LLM) to evaluate the initial 63 items, comparing its assessments with those of the human judges. Interestingly, the AI provided also detailed explanations for its categorisation. Results showed no significant difference between AI and human judges for 35 items, minor differences for 23 items, and major differences for 5 items. This experiment demonstrates the potential of integrating AI with traditional psychometric methods to enhance the development of attitude measurement scales.

## 1. Introduction

For nearly a century, social researchers and psychologists employing psychometric scales have investigated attitude measurement, primarily deliberating between Thurstone scales of *equal appearing intervals* and Likert's *summated rating scales* (Edwards and Kenney, 1946; Nunnally and Bernstein, 1994; Chernyshenko et al., 2007; Drasgow et al. 2010; Willits et al., 2016). Thurstone scales declined in popularity due to criticism regarding the time-consuming process required to collect the judges' opinions on the initial set of items, which constituted the primary argument against their use. However, with recent technological advancements, the challenges previously faced when using a group of judges to construct Thurstone scales have been overcome. Consequently, the perceived simplicity of Likert scales is no longer a significant advantage, as Likert scales themselves exhibit their own methodological challenges (Symeonaki et al. 2015; Drasgow et al. 2010; Carifio and Perla, 2007). In this study, we conduct a methodological experiment to develop a Thurstone scale with equally appearing intervals for assessing attitudes toward individuals living with AIDS (Kazani et al. 2024). An electronic questionnaire was distributed to a group of seventy-five (75) individuals, composed of undergraduate, postgraduate, and PhD students from various disciplines, namely social policy, law, medicine, and computer engineering, as well as established researchers in social research, who acted as the group of judges. Their responses were easily collected and subjected to statistical analysis. Additionally, and this represents the main benefit of the proposed method, an Artificial Intelligence (AI) Large Language Model (LLM) was enlisted to provide its assessment of the initial set of 63 items, and its responses were compared with those of the judges. To our knowledge, this is the first time where an LLM has been utilised to assist in the construction of an attitude scale. The model also offered a comprehensive explanation for its selection, elucidating why each item was assigned to its specific category. The analysis revealed that in 35 out of the 63 initial items, there is no difference or only a very slight difference between the judges' opinions and the model's. Some differences are detected in 23 out of the 63 initial items, while 5 items exhibit a major difference in relation to the categorisation.

The structure of this paper is organised as follows: Section 2 offers a detailed overview of the fundamental principles and processes involved in constructing a Thurstone scale, including historical context and methodological considerations. Section 3 presents the findings from our methodological experiment, highlighting the role of the AI Large Language Model (LLM) in evaluating and categorising attitude items, and comparing these results with those obtained from human judges. Finally, Section 4 discusses the implications of our findings, summarises the key conclusions, and suggests potential directions for future research in the application of AI in psychometric scale development.

## 2. Methodology of the development of an attitude scale using a Large Language Model

### 2.1 Description of Thurstone scale methodology

Thurstone scaling, also known as the method of equal-appearing intervals, was developed by Louis L. Thurstone (Thurstone 1927; 1928; 1929) to create quantitative unidimensional attitude measurement instruments. Unlike earlier methods, Thurstone scaling produces an interval scale, enabling the use of parametric statistics for analysis. Despite being time-consuming and complex it has significantly contributed to the development of rating scales (Saucer, 2010).

The idea behind the equal-appearing intervals technique is to develop a unidimensional scale by selecting a number of agree-disagree statements in order to measure the attitude towards e.g. a specific group of individuals. The method begins by gathering a sufficiently large pool of opinion statements on a specific issue. After reviewing the statements, they are given to a group of judges who are instructed to categorise them on a scale from 1, indicating the most negative attitude, to 11, indicating the most positive attitude regarding the issue. It is important that the judges do not express agreement or disagreement with the statements; instead, they should rate the statements based on how effectively they reveal a positive or negative attitude. As a third step, the median and the corresponding measure of variability, the interquartile range, are calculated for each statement. The statements are then presented in a table, sorted in ascending order by median and by the respective interquartile range. The final statements are selected to cover all categories, from 1 to 11, and are chosen based on the smallest interquartile range, indicating the highest agreement among the judges. In case of a tie between statements, many textbooks recommend a "rule of thumb" where the final decision is left to the researcher.

Comparability of Thurstone and Likert scales (Likert, 1932) involve mainly the following: (i) the time involved in the construction of the scale, (ii) the need of the judges and their influence in the final scoring, and (iii) the reliability of the scales produced. Apparently, Thurstone scales gradually fell out of use due to the labour-intensive nature and efforts of collecting and analysing judges' opinions on the initial items, and the concern that these opinions might bias the scale's construction. However, our study demonstrates that the effort required today is significantly reduced compared to what was the case a century ago. Moreover, by incorporating a Large Language Model (LLM) alongside human judges, we effectively minimise the influence and subjectivity associated with the judges' group, making the construction process more efficient and objective.

### 2.2 Description of the suggested methodology

The proposed methodology integrates traditional psychometric techniques with advanced artificial intelligence (AI) tools to enhance efficiency and objectivity. The process begins with the selecting the initial set of items designed to measure attitudes towards individuals living with AIDS[1]. These items are then evaluated by a group of judges, including undergraduate, postgraduate, and PhD students from various disciplines as earlier mentioned. An electronic questionnaire was distributed to this group and their responses were easily collected and subjected to statistical analysis. Overall, data collection and descriptive statistical analysis took approximately one month.
Having statistically analysed the items, we utilise a Large Language Model (LLM) to assist in the item evaluation process. The LLM assesses the same set of items and provides categorisations based on sentiment and relevance, accompanied by detailed explanations for each categorisation. This AI-driven approach allows for a comparative analysis between the AI's assessments and the judges' evaluations.
The responses from both the judges and the output of the LLM model, namely OpenAI (2023), when prompted with "Categorise the following statement from 1=most negative to 11=most positive when measuring attitudes towards individuals with AIDS", are compared to determine the degree of alignment and identify any discrepancies. Items with significant differences in categorisation should be further scrutinised and refined to ensure accuracy and validity.
By combining human judgment with AI capabilities, this methodology not only reduces the effort required in scale construction but also addresses concerns about subjectivity and bias. The result is a more robust and reliable Thurstone scale that leverages modern technological advancements to improve traditional psychometric practices.

[1] Trochim, W. [Online]. Thurstone scaling [cited 2024 May25]; Available from: URL: https://conjointly.com/kb/thurstone-scaling/

## 3. Results

The evaluations from both the judges and the LLM are juxtaposed and illustrated in the subsequent tables to ascertain the level of congruence and pinpoint any inconsistencies. The examination unveiled that among the 63 initial items, there is either no variance or merely a minimal one in 35 instances between the judges' assessments and those of the model. Small differences were identified in 23 of the 63 initial items, while 5 items displayed significant disparities in their categorisation.
The items that could be selected for the final scale are best considered to be those where the judges and the LLM model "agree" about the category since this agreement indicates a high level of reliability and validity. When both human experts (judges) and the language model consistently categorise an item the same way, it suggests that the item is clearly defined and understood, minimising ambiguity. This consensus helps ensure that the scale accurately measures what it is intended to measure, thereby enhancing the overall credibility and utility of the scale.
We now list the statements that could be considered for the final scale per category where there is agreement between the judges and the model. An exception exists manly in the statement that represents the 6$^{th}$ category.

**Category 1**

1. AIDS is the wrath of God (Q=0)
2. People who contract AIDS deserve it (Q=0)
3. People with AIDS are bad (Q=0)
4. People with AIDS deserve what they got (Q=0)
5. AIDS is good because it will help control the population (Q=0)
6. Everyone affected with AIDS deserves it due to their lifestyle (Q=0)

Any selection from these six statements is acceptable for representing the 1$^{st}$ category, as the judges and the model concur, and the interquartile rate is zero (Q=0), indicating complete agreement among the judges.

**Category 2**

1. AIDS is something the other guy gets (Q=2)
2. Because I don't live in a big city, the threat of AIDS is very small (Q=2)
3. AIDS is a omnipresent, ruthless killer that lurks around dark alleys, silently waiting for naive victims to wander passed so that it might pounce (Q=2)
4. AIDS will never happen to me (Q=3)
5. Children cannot catch AIDS (Q=4)
6. Bringing AIDS into my family would be the worst thing I could do (Q=4)
7. The AIDS quilt is an emotional reminder to remember those who did not deserve to die painfully or in vain (Q=4)

The best statements to denote the 2$^{nd}$ category are among the first three, where Q shows the smallest value.

**Category 3**

1. Very few people have AIDS, so it's unlikely that I'll ever come into contact with a sufferer (Q=3)
2. By the time I would get sick from AIDS, there will be a cure (Q=3)
3. I can't get AIDS if I'm in a monogamous relationship (Q=4)

The same applies to Category 3, where the best statements are listed among the first two, with Q showing the smallest value.

**Category 4**

1. AIDS is becoming more a problem for heterosexual women and their offsprings than IV drug users or homosexuals (Q=3) (ChatGPT=3)

The only suitable choice for Category 4 is the aforementioned question, where the model assigns a category one level lower and Q exhibits the smallest value.

**Category 5**

1. The number of individuals with AIDS in Hollywood is higher than the general public thinks (Q=3)

Category 5 is represented by a statement where the judges and the model "agree" and Q=3.

**Category 6**

1. **It's easy to get AIDS (Q=3, ChatGPT=3)**

This discrepancy underscores a difficulty in finding an appropriate option for the 6th category when there is inconsistency between human judgment and the model's assessment.

**Category 7**

1. A cure for AIDS is on the horizon (Q=4)

Category 7 is represented by a statement where the judges and the model are in "agreement", with Q equal to 4.

**Category 8**

1. It is not the AIDS virus that kills people, it is complications from other illnesses (because the immune system isn't functioning) that cause death (Q=5)
2. I know enough about the spread of the disease that I would have no problem working in a health care setting with patients with AIDS (Q=6)

The first of the suggested statements should be favoured due to its smaller Q value.

**Category 9**

1. Aids affects us all (Q=5, ChatGPT=9-10)

The sole option is the one mentioned above, where the group of judges and the model nearly concur.

**Category 10**

1. Someone with AIDS could be just like me (Q=2)
2. AIDS does not discriminate (Q=3)

We opt for the initial suggested statement, as it exhibits a smaller Q value.

**Category 11**

1. I treat everyone the same, regardless of whether or not they have AIDS (Q=1, ChatGPT=9)

For the final category, the optimal choice would be the suggested question where the group of judges appear to agree (Q=1), albeit the model categorises it in a lower category.

**AIDS is the wrath of God**

| **Judges** | | | **ChatGPT** |
|---|---|---|---|
| Median | Interquartile Range | Bar Chart | Category and explanation |
| **1** | **0** | 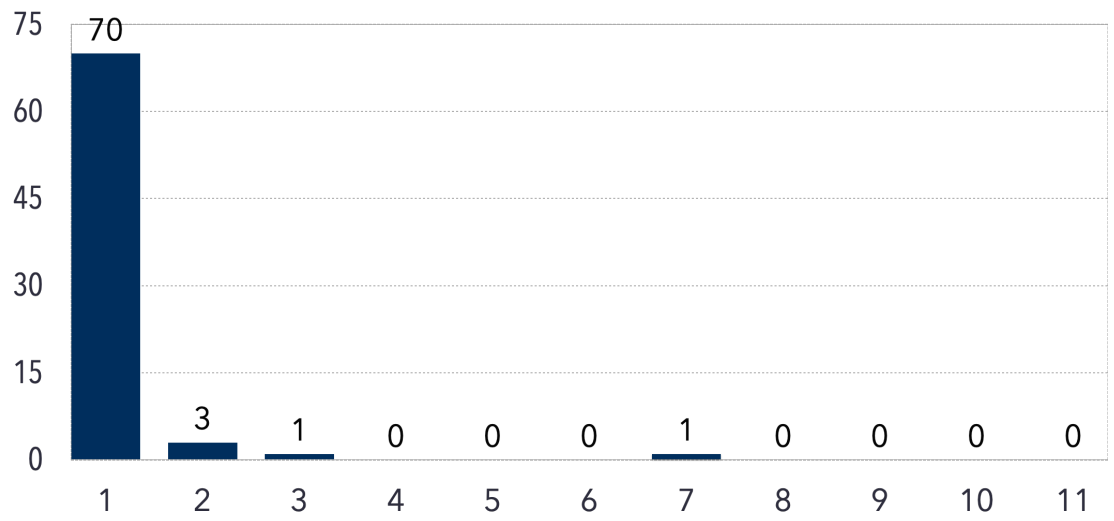 | **1**<br>This statement would be rated as 1 on the scale, indicating the most negative judgment. This statement reflects a harmful and stigmatizing belief that associates HIV/AIDS with divine punishment or retribution. Such beliefs contribute to the stigma surrounding HIV/AIDS and can lead to discrimination against individuals living with the virus. There is no scientific basis for this assertion, as HIV/AIDS is a medical condition caused by a virus, not a punishment from a higher power. This belief perpetuates harmful stereotypes and can hinder efforts to promote understanding, compassion, and effective prevention and treatment of HIV/AIDS. Therefore, this statement would also be rated low on the scale, as it represents a deeply negative and stigmatizing view of HIV/AIDS. |

**AIDS is a death sentence**

| **Judges** | | | **ChatGPT** |
|---|---|---|---|
| Median | Interquartile Range | Bar Chart | Category and explanation |
| **1** | **0** | 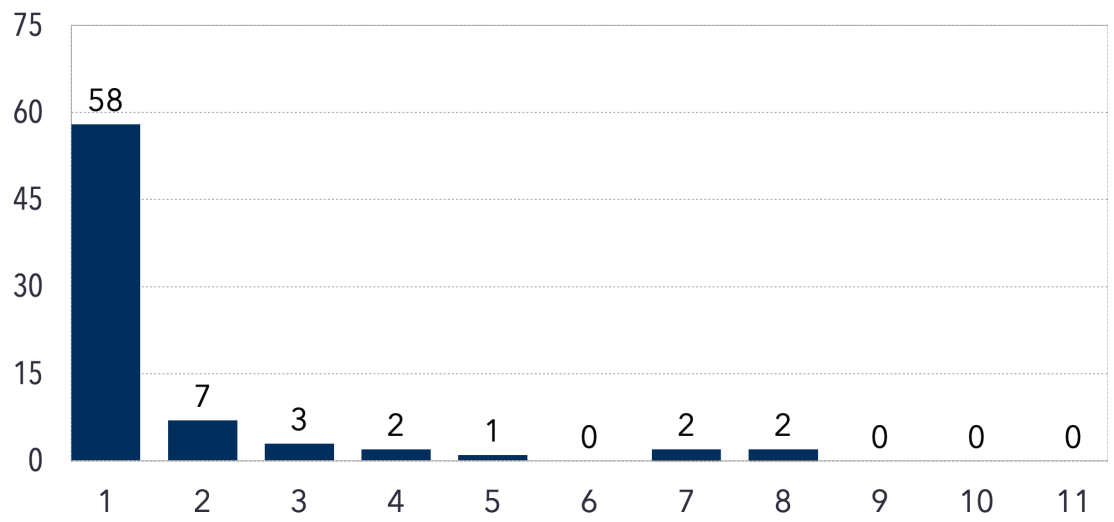 | **2**<br>While the statement does recognize the serious nature of AIDS, it is outdated and overly negative given the medical advancements and current understanding of HIV/AIDS. Therefore, it warrants a rating of 2 on the scale. |

**AIDS distracts from other diseases that deserve our attention more**

| **Judges** | | | **ChatGPT** |
|---|---|---|---|
| Median | Interquartile Range | Bar Chart | Category and explanation |
| **1** | **0** | 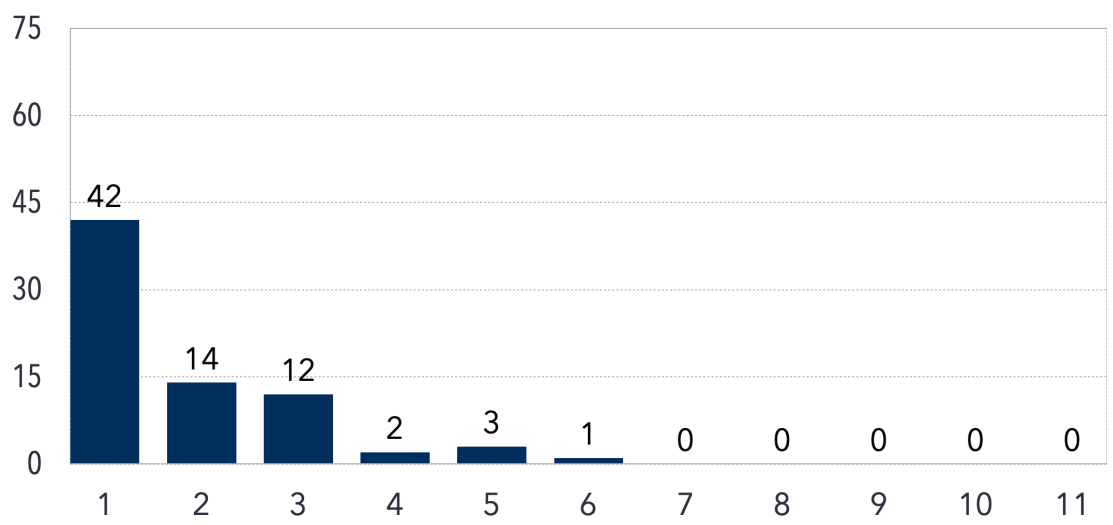 | **3**<br>This statement suggests a belief that resources and attention allocated to AIDS could be better directed towards other diseases perceived as more deserving. It implies a prioritization of healthcare resources and research efforts away from AIDS. This viewpoint may be perceived negatively by some, as it could be seen as diminishing the importance of addressing AIDS as a global health issue. However, others may view it as a pragmatic consideration of resource allocation in the context of competing health priorities. Overall, it carries a somewhat negative sentiment, potentially warranting a rating around 3 on the scale. |

**People who contract AIDS deserve it**

| **Judges** | | | **ChatGPT** |
|---|---|---|---|
| Median | Interquartile Range | Bar Chart | Category and explanation |
| **1** | **0** | 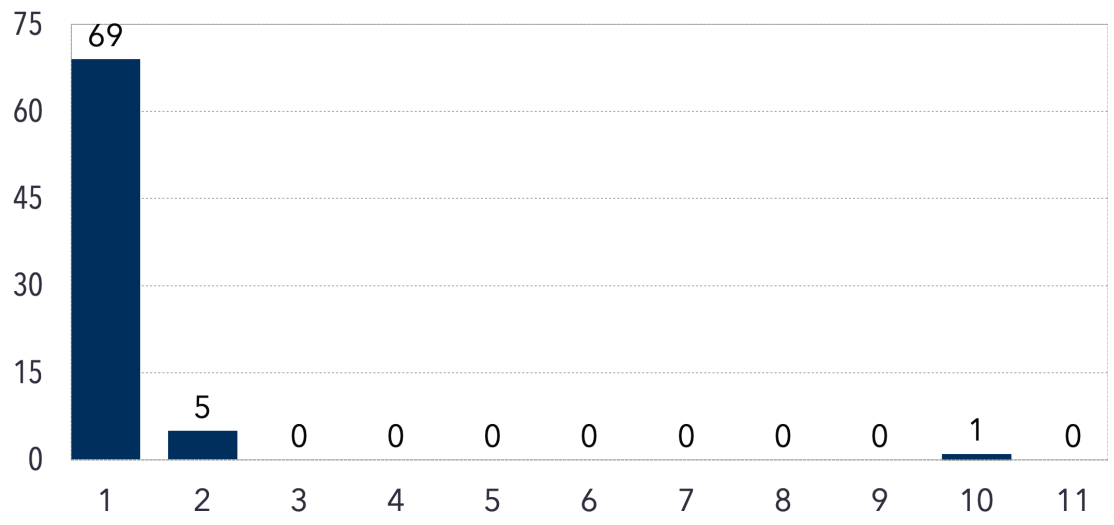 | **1**<br>This statement would likely be rated around 1. It expresses a highly stigmatizing and discriminatory attitude toward people living with HIV/AIDS, suggesting that they somehow deserve their condition. Such beliefs are harmful and unfounded, as HIV/AIDS can affect anyone, regardless of their actions or behaviors. |

**Bad people get AIDS and since I am a good person I will never get AIDS**

| **Judges** | | | **ChatGPT** |
|---|---|---|---|
| Median | Interquartile Range | Bar Chart | Category and explanation |
| **1** | **0** | 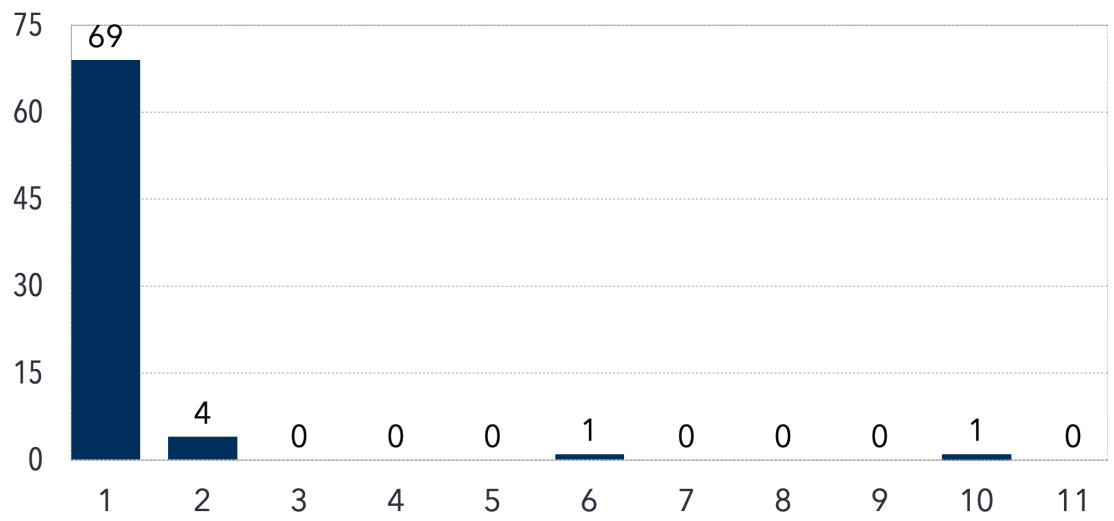 | **1-2**<br>This statement reflects a deeply negative and judgmental attitude towards individuals with AIDS, attributing the condition to moral character rather than recognizing it as a medical issue. It also implies a false sense of security for the speaker based on their perceived moral superiority. This attitude contributes to stigma and discrimination against people living with HIV/AIDS. Hence, it would be rated very low on the scale, likely a 1 or 2, indicating a highly negative sentiment. |

**People with AIDS are bad**

| **Judges** | | | **ChatGPT** |
|---|---|---|---|
| Median | Interquartile Range | Bar Chart | Category and explanation |
| **1** | **0** | 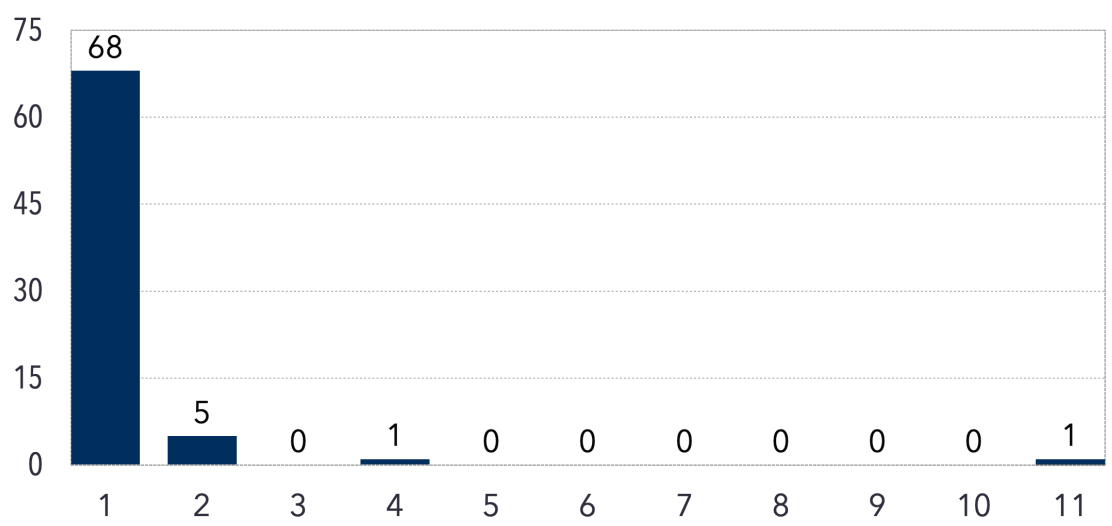 | **1**<br>This statement would be rated as 1, representing the most negative judgment. It conveys a stigmatizing and discriminatory attitude towards individuals with AIDS, unfairly characterizing them as morally flawed or unworthy. |

**People with AIDS deserve what they got**

| **Judges** | | | **ChatGPT** |
|---|---|---|---|
| Median | Interquartile Range | Bar Chart | Category and explanation |
| **1** | **0** | 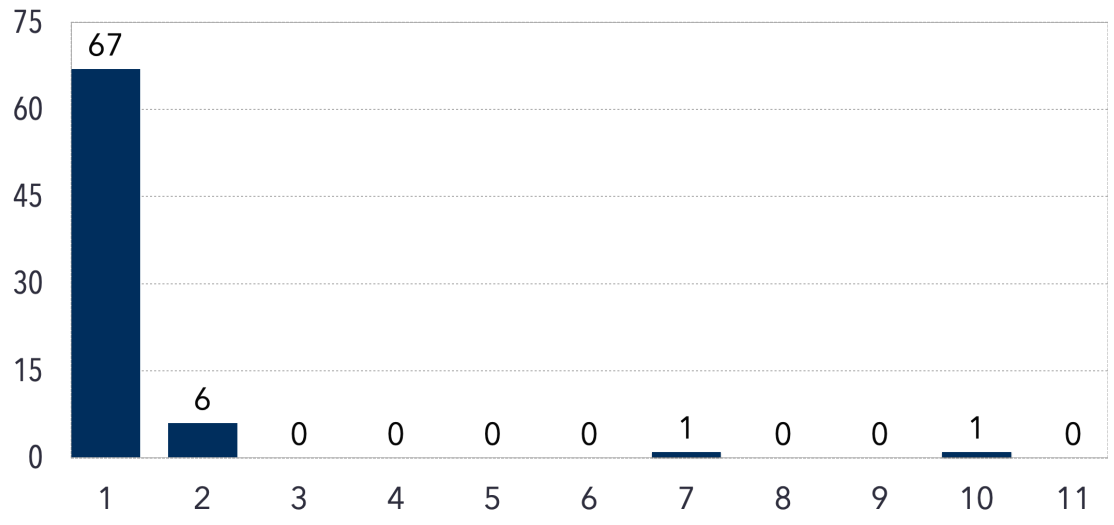 | **1**<br>This statement expresses a negative attitude towards people with AIDS, suggesting blame or judgment for their condition. It implies a lack of empathy and understanding of the complexities surrounding HIV transmission and the challenges faced by individuals living with AIDS. Therefore, it would be rated very low on the scale, likely a 1, indicating a highly negative sentiment towards individuals with AIDS. |

**People with AIDS do not need or deserve our help**

| **Judges** | | | **ChatGPT** |
|---|---|---|---|
| Median | Interquartile Range | Bar Chart | Category and explanation |
| **1** | **0** | 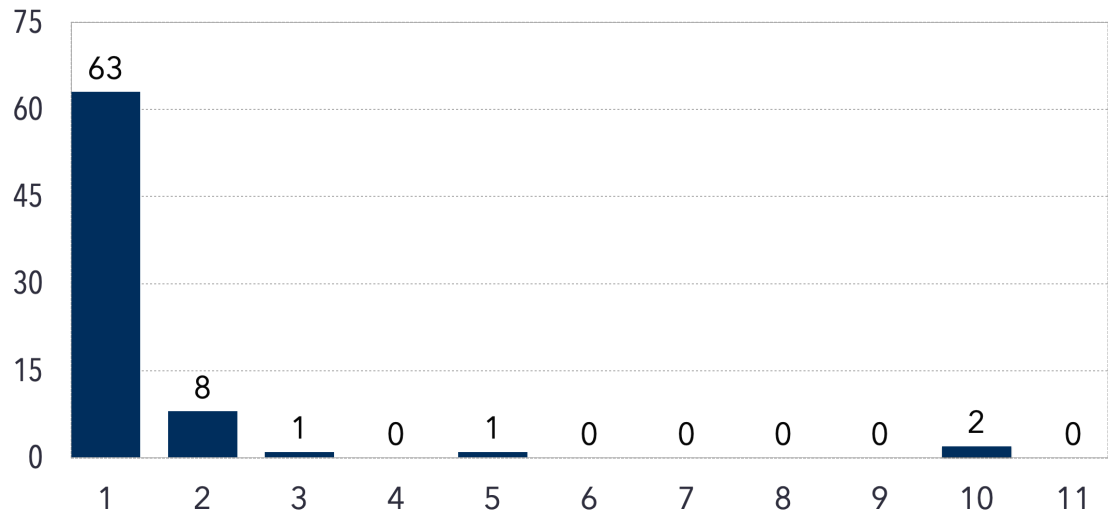 | **2**<br>This statement reflects a negative and callous attitude towards individuals with AIDS, suggesting a lack of empathy or support for them. It implies that those affected by AIDS should not receive assistance or compassion from others. This sentiment is quite harsh and unsympathetic, warranting a low rating on the scale, perhaps around 2. |

**If my brother caught AIDS I'd never talk to him again**

| **Judges** | | | **ChatGPT** |
|---|---|---|---|
| Median | Interquartile Range | Bar Chart | Category and explanation |
| **1** | **0** | 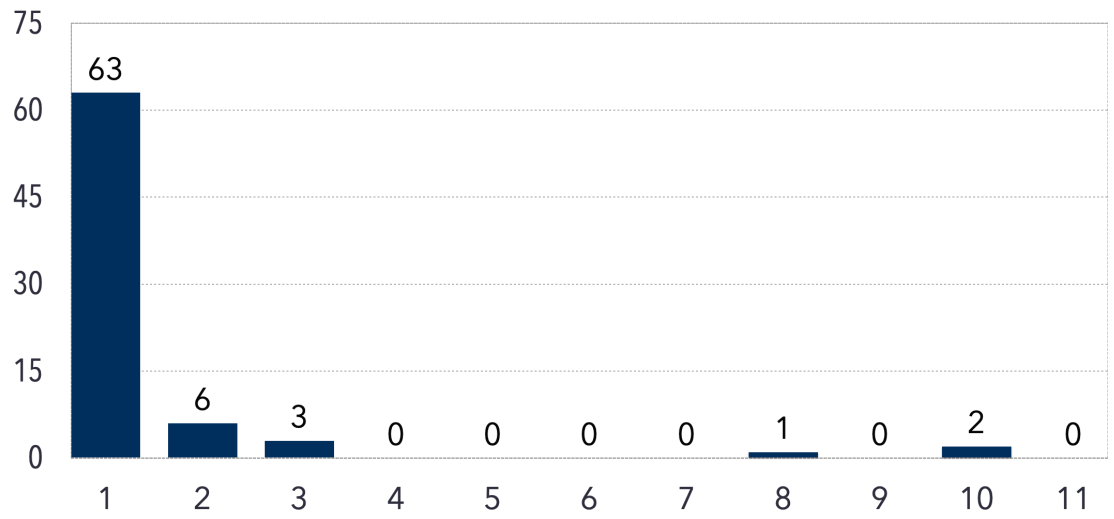 | **1-2**<br>This statement is extremely negative and reflects a harmful, stigmatizing attitude towards individuals living with HIV/AIDS. It warrants a rating of 1 on the scale due to its potential to cause significant emotional harm and perpetuate misinformation and prejudice. Family support is crucial for individuals living with chronic illnesses, including HIV/AIDS. Rejection from family members can have devastating effects. Showing compassion and empathy towards a loved one facing a health challenge is essential for their well-being and recovery. |

**AIDS is good because it will help control the population**

| Judges | | | ChatGPT |
|---|---|---|---|
| Median | Interquartile Range | Bar Chart | Category and explanation |
| **1** | **0** | 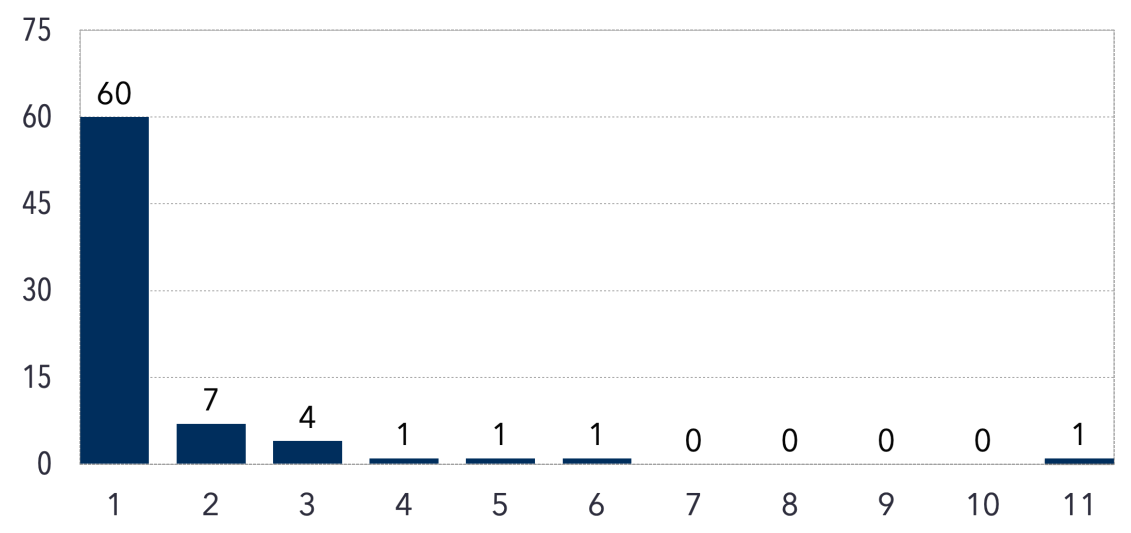 | **1**<br>This statement is extremely negative and unethical. It suggests a callous disregard for human life and suffering, as well as advocating for a harmful and discriminatory belief. Therefore, it would be rated as a 1. |

**Everyone affected with AIDS deserves it due to their lifestyle**

| Judges | | | ChatGPT |
|---|---|---|---|
| Median | Interquartile Range | Bar Chart | Category and explanation |
| **1** | **0** | 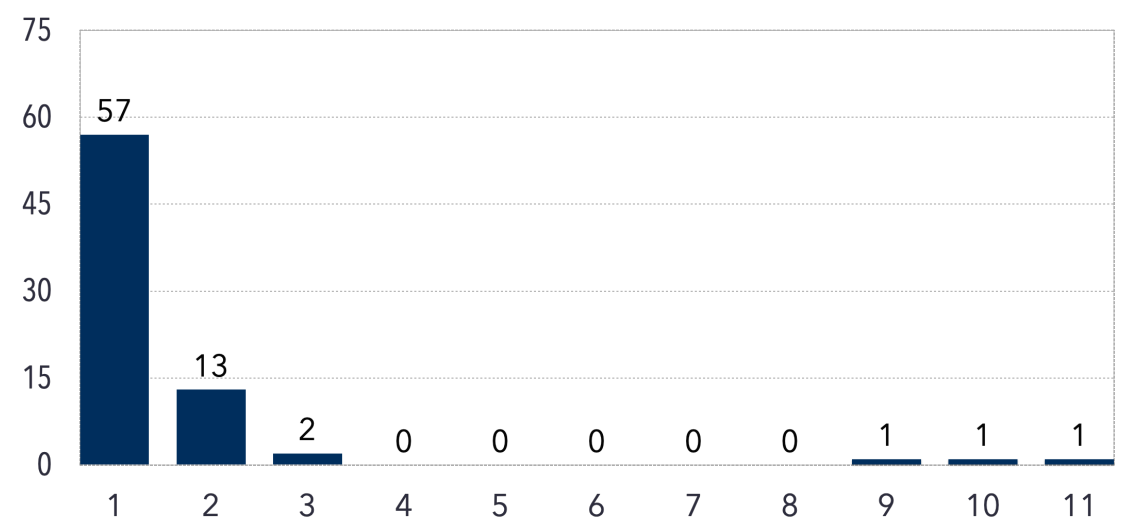 | **1**<br>This statement conveys a judgmental and stigmatizing attitude towards individuals with AIDS, suggesting that they are solely responsible for their condition because of their lifestyle choices. Such a perspective is highly negative and contributes to the stigma surrounding AIDS. Therefore, it would be rated very negatively, perhaps around 1, for its lack of empathy and understanding towards those affected by the disease. |

**Women don't get AIDS**

| Judges | | | ChatGPT |
|---|---|---|---|
| Median | Interquartile Range | Bar Chart | Category and explanation |
| **1** | **1** | 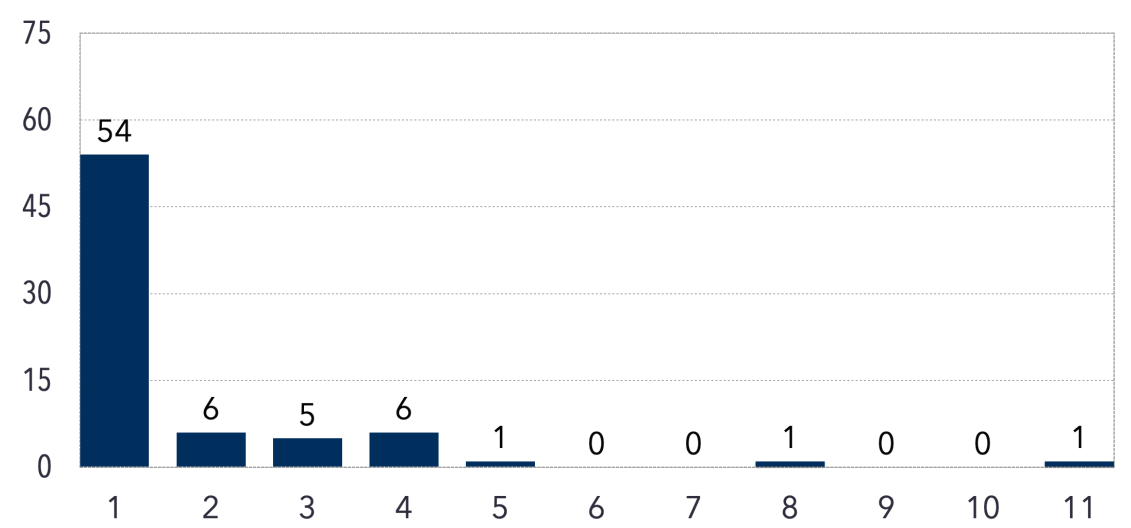 | **2**<br>This statement would likely rank around a 2 on the scale. It reflects a significant misconception about AIDS transmission, as women can indeed contract HIV, the virus that causes AIDS, through various means including sexual contact, sharing needles, and childbirth. |

**People with AIDS should be quarantined to protect the rest of society**

| Judges | | | ChatGPT |
|---|---|---|---|
| Median | Interquartile Range | Bar Chart | Category and explanation |
| **1** | **1** | 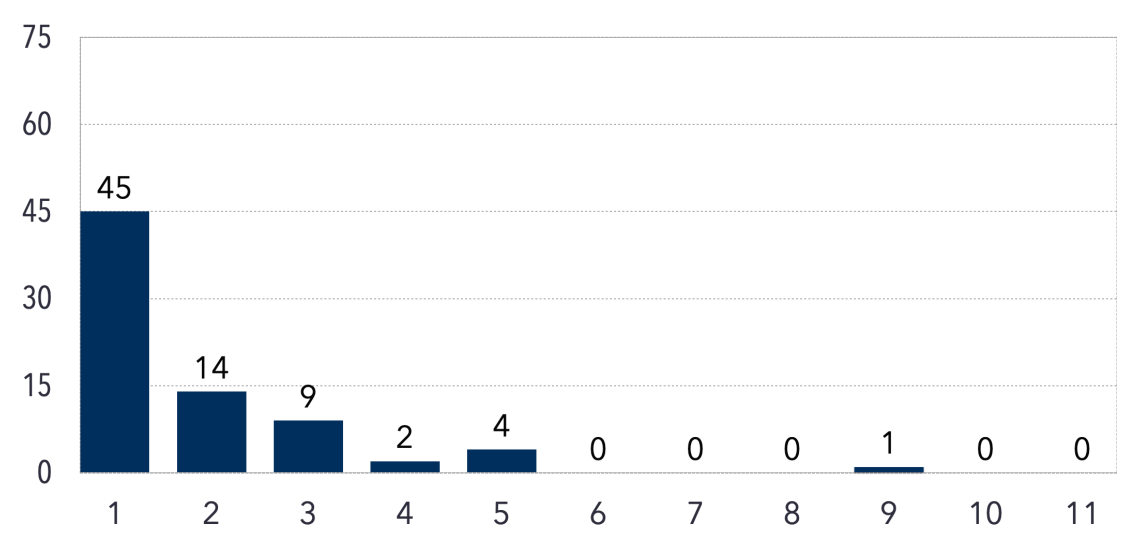 | **2**<br>This statement suggests a discriminatory attitude towards people with AIDS, advocating for their isolation from society. Such a viewpoint is harmful and stigmatizing, as it overlooks the rights and dignity of individuals living with AIDS. Therefore, it would be rated quite negatively, perhaps around 2, as it promotes segregation and discrimination. |

**AIDS is spread the same way colds are**

| **Judges** | | | **ChatGPT** |
|---|---|---|---|
| Median | Interquartile Range | Bar Chart | Category and explanation |
| **1** | **2** | 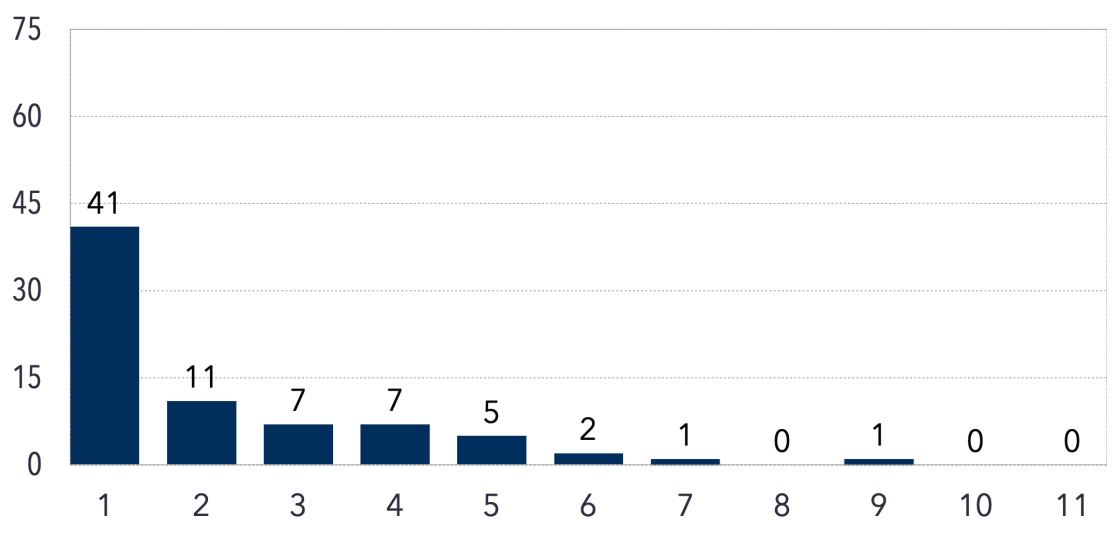 | **2**<br>On the scale, this statement would rank very low, around 2, due to its significant inaccuracy regarding HIV transmission. |

**People get AIDS by engaging in immoral behaviour**

| **Judges** | | | **ChatGPT** |
|---|---|---|---|
| Median | Interquartile Range | Bar Chart | Category and explanation |
| **1** | **2** | 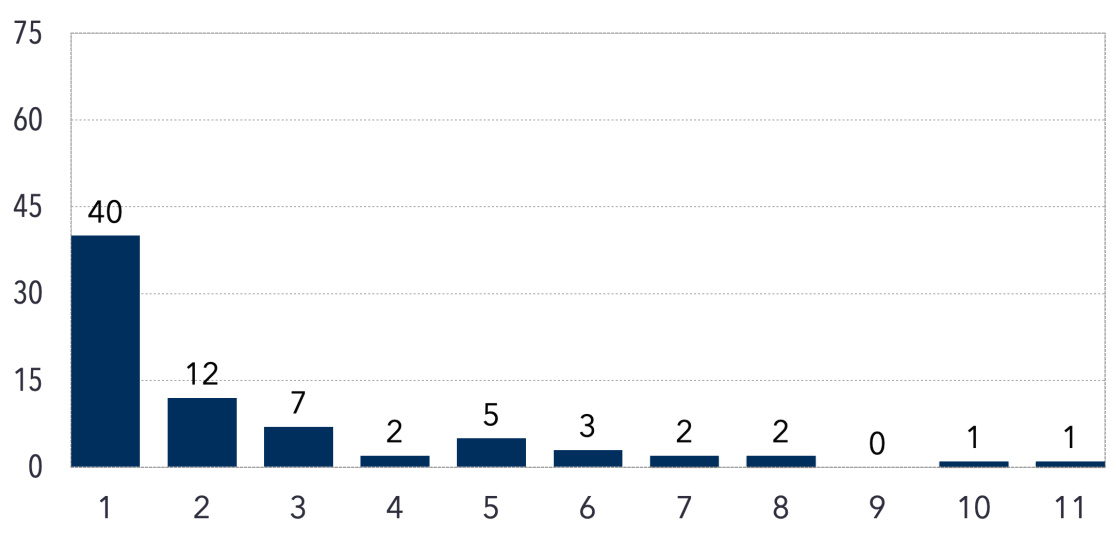 | **1-2**<br>This statement implies a moral judgment about individuals who contract AIDS, suggesting that it is a result of immoral behavior. Such a belief contributes to stigmatization and discrimination against people living with HIV/AIDS, perpetuating harmful stereotypes and misinformation. It overlooks the various ways in which HIV can be transmitted, including through unprotected sex, needle sharing, and mother-to-child transmission, which can occur regardless of moral behavior. Additionally, labeling certain behaviors as "immoral" implies a value judgment that is subjective and may vary across cultures and societies. Overall, this statement reflects a negative and judgmental attitude towards individuals affected by AIDS, hence the low rating on the scale. |

**AIDS is spread through the air**

| **Judges** | | | **ChatGPT** |
|---|---|---|---|
| Median | Interquartile Range | Bar Chart | Category and explanation |
| **1** | **4** | 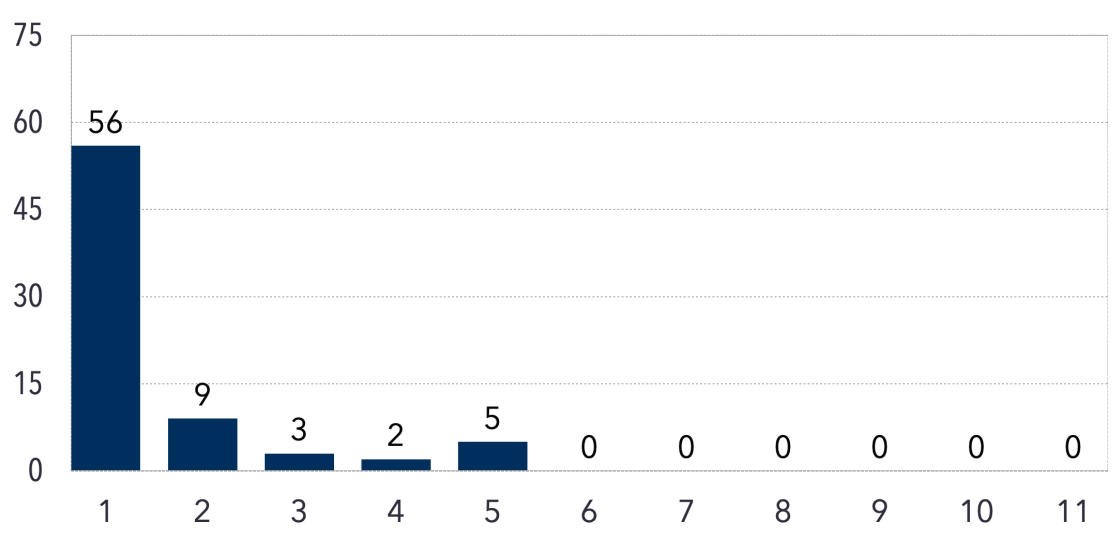 | **2**<br>This statement is largely negative for those with AIDS, as it contributes to fear and misunderstanding surrounding the virus, potentially leading to increased stigma and discrimination against people living with HIV/AIDS. Therefore, it would typically be rated low on the scale, around 2. |

**Category 2**

**AIDS is something the other guy gets**

| **Judges** | | | **ChatGPT** |
|---|---|---|---|
| Median | Interquartile Range | Bar Chart | Category and explanation |
| **2** | **2** | 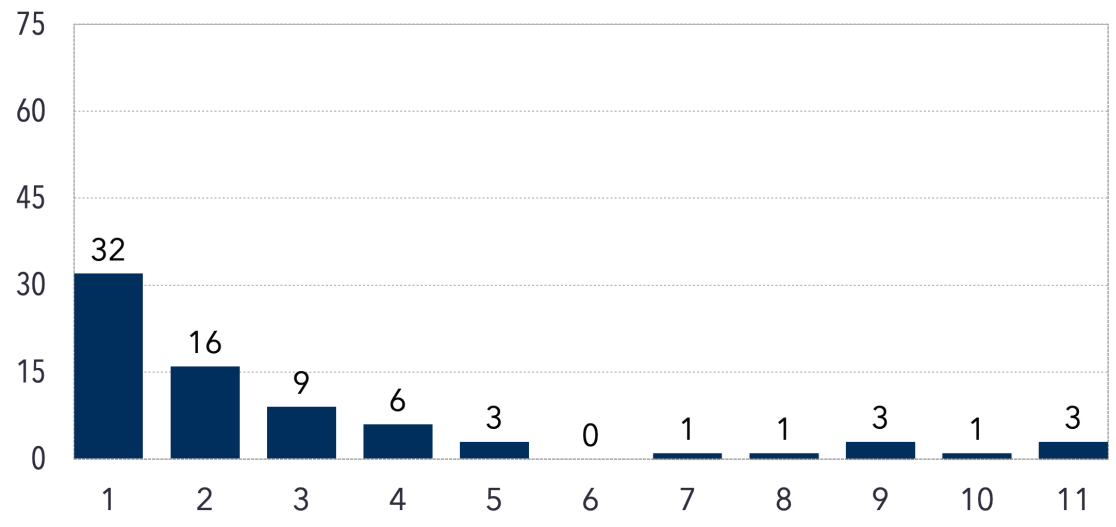 | **2**<br>This statement could be rated quite negatively, maybe around 2. It reflects a sense of denial or distancing from the possibility of contracting AIDS oneself, which can indicate a lack of awareness or empathy towards the seriousness of the disease. |

**I don't care if I get AIDS because researchers will soon find a cure for it**

| **Judges** | | | **ChatGPT** |
|---|---|---|---|
| Median | Interquartile Range | Bar Chart | Category and explanation |
| **2** | **2** | 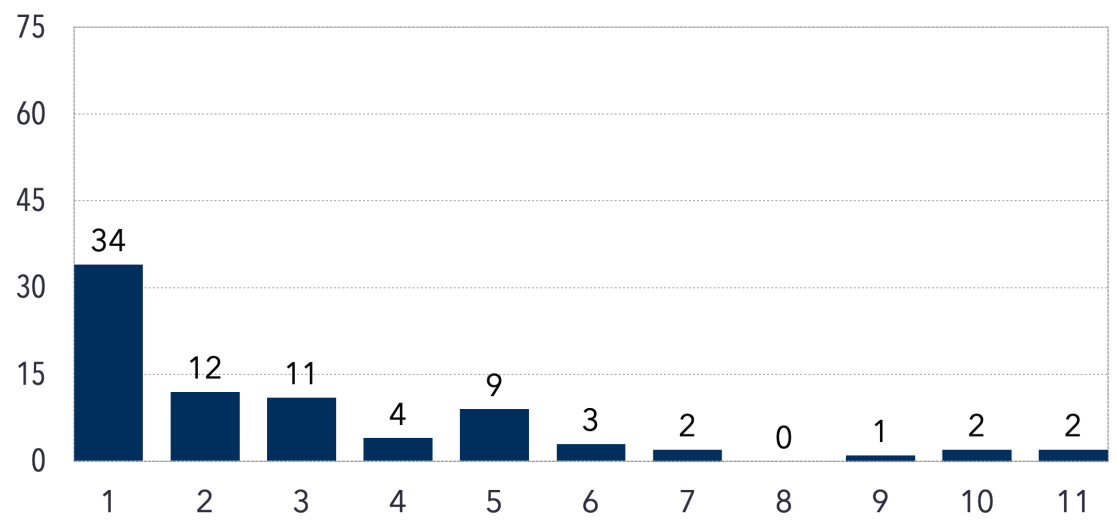 | **3**<br>While the statement is not as overtly stigmatizing or harmful to others as some other statements, it reflects a dangerously naive and irresponsible mindset towards a serious health issue. It undermines the importance of prevention and misrepresents the current state of medical research on HIV/AIDS. Therefore, it warrants a rating of 3 on the scale due to its potential to encourage risky behavior and spread misinformation. |

**If you get AIDS you will die within a year**

| **Judges** | | | **ChatGPT** |
|---|---|---|---|
| Median | Interquartile Range | Bar Chart | Category and explanation |
| **2** | **2** | 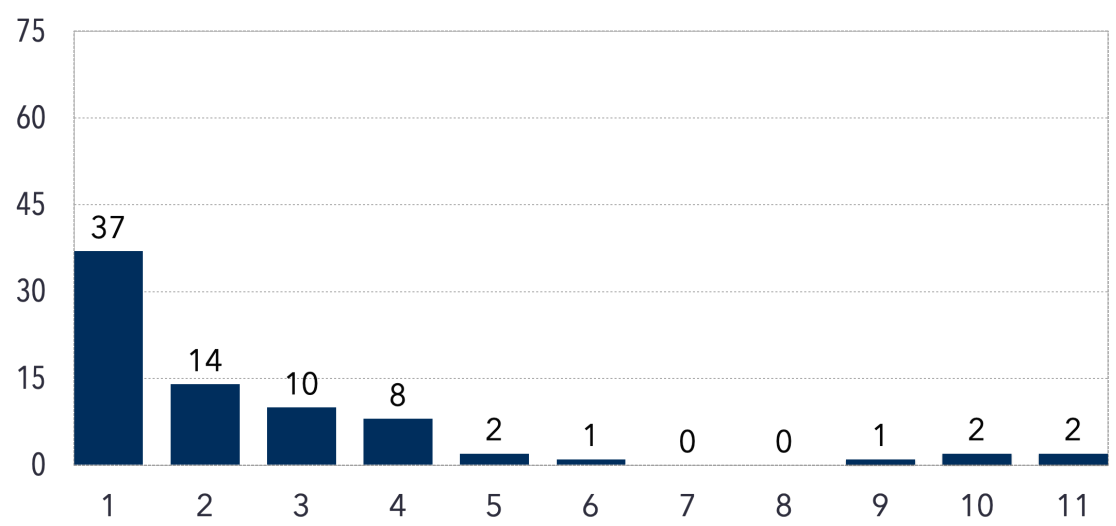 | **1**<br>This statement conveys a very negative and inaccurate belief about AIDS, suggesting that contracting the disease inevitably leads to death within a year. Such a belief perpetuates stigma and misinformation surrounding HIV/AIDS, disregarding the advancements in medical treatment and the possibility of living a long and fulfilling life with proper care. Therefore, it would be rated very low on the scale, likely a 1, indicating a highly negative sentiment. |

**Anybody with AIDS is either gay or a junkie**

| **Judges** | | | **ChatGPT** |
|---|---|---|---|
| Median | Interquartile Range | Bar Chart | Category and explanation |
| **2** | **2** | 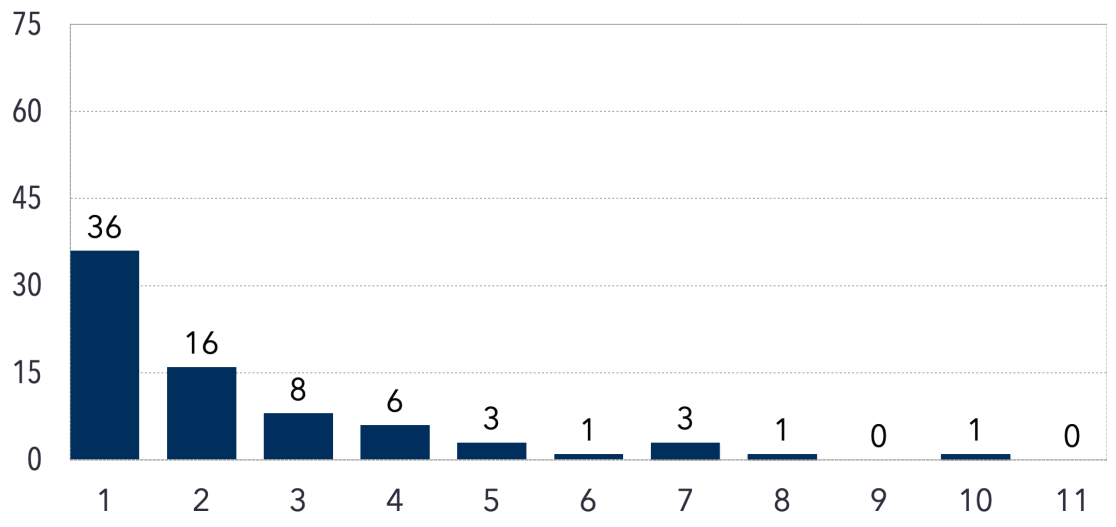 | **1**<br>This statement would also be rated as 1 on the scale, indicating the most negative judgment. It reflects a deeply stigmatizing and discriminatory attitude towards individuals with AIDS, perpetuating harmful stereotypes and misconceptions. |

**Because I don’t live in a big city, the threat of AIDS is very small**

| **Judges** | | | **ChatGPT** |
|---|---|---|---|
| Median | Interquartile Range | Bar Chart | Category and explanation |
| **2** | **2** | 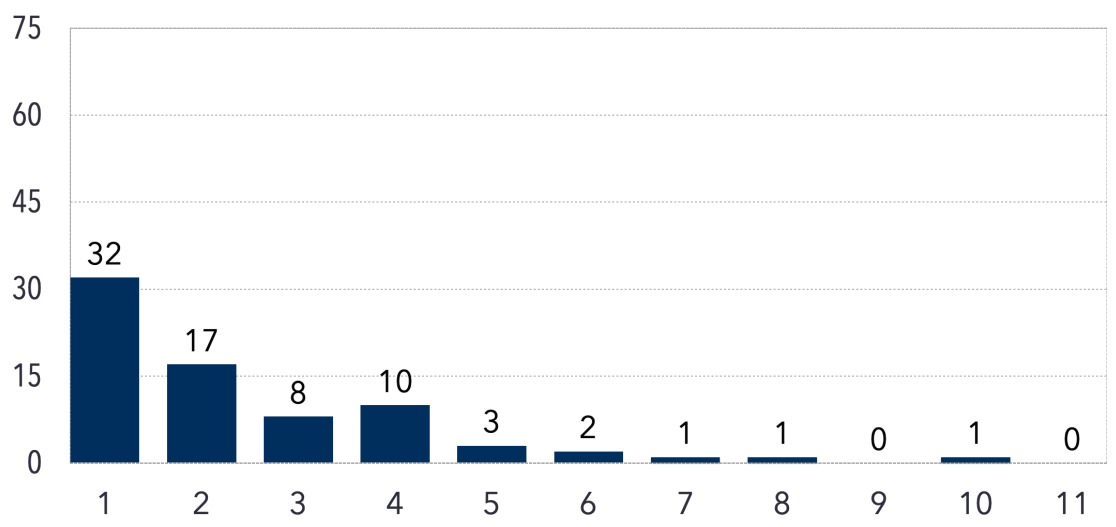 | **2**<br>The statement is misleading and fosters a dangerous sense of complacency about HIV/AIDS risk based on geographic location. It downplays the universal nature of HIV risk factors and the importance of preventive measures for everyone. Therefore, it warrants a rating of 2 on the scale due to its potential to spread misinformation and increase risk behaviors across different populations. |

**AIDS is a omnipresent, ruthless killer that lurks around dark alleys, silently waiting for naive victims to wander passed so that it might pounce**

| **Judges** | | | **ChatGPT** |
|---|---|---|---|
| Median | Interquartile Range | Bar Chart | Category and explanation |
| **2** | **2** | 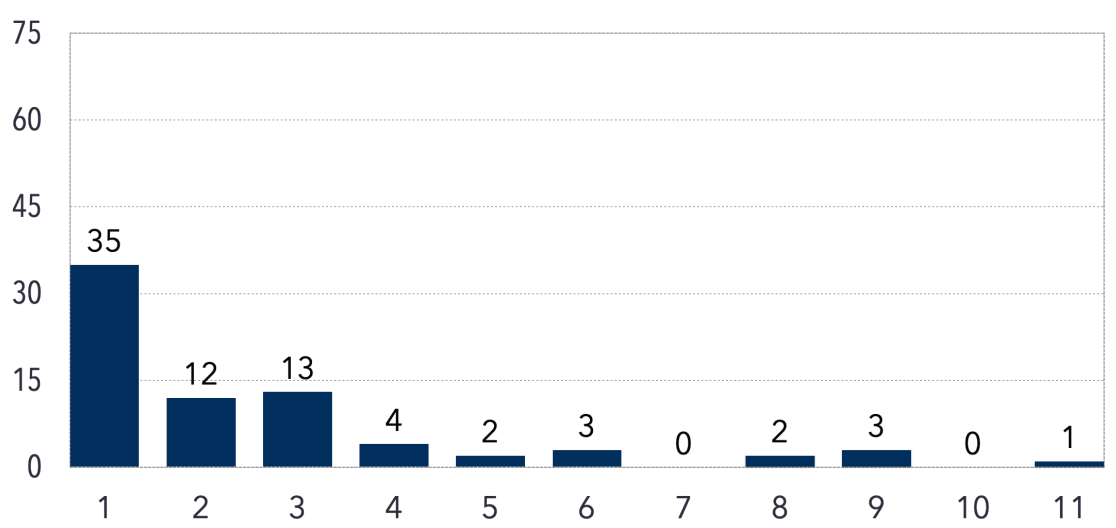 | **2**<br>This statement portrays a highly negative and fearful perception of AIDS, depicting it as a menacing and omnipresent threat. It emphasizes the idea of AIDS as a deadly predator, waiting to strike unsuspecting victims. On the scale, it would likely be rated towards the lower end, perhaps around 2, indicating a strongly negative view. |

## AIDS costs the public too much

| Judges | | | ChatGPT |
|---|---|---|---|
| Median | Interquartile Range | Bar Chart | Category and explanation |
| **2** | **3** | 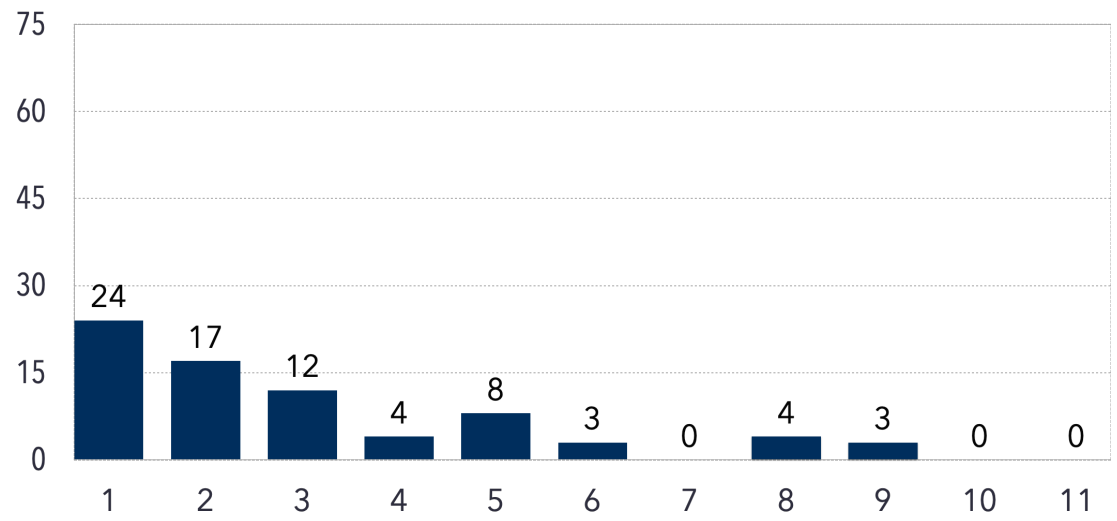 | **3**<br>This statement would likely be rated towards the negative end of the scale, perhaps around 3. It suggests a concern about the financial burden of AIDS on the public, which can imply a lack of empathy or understanding towards those affected by the disease. |

## AIDS will never happen to me

| Judges | | | ChatGPT |
|---|---|---|---|
| Median | Interquartile Range | Bar Chart | Category and explanation |
| **2** | **3** | 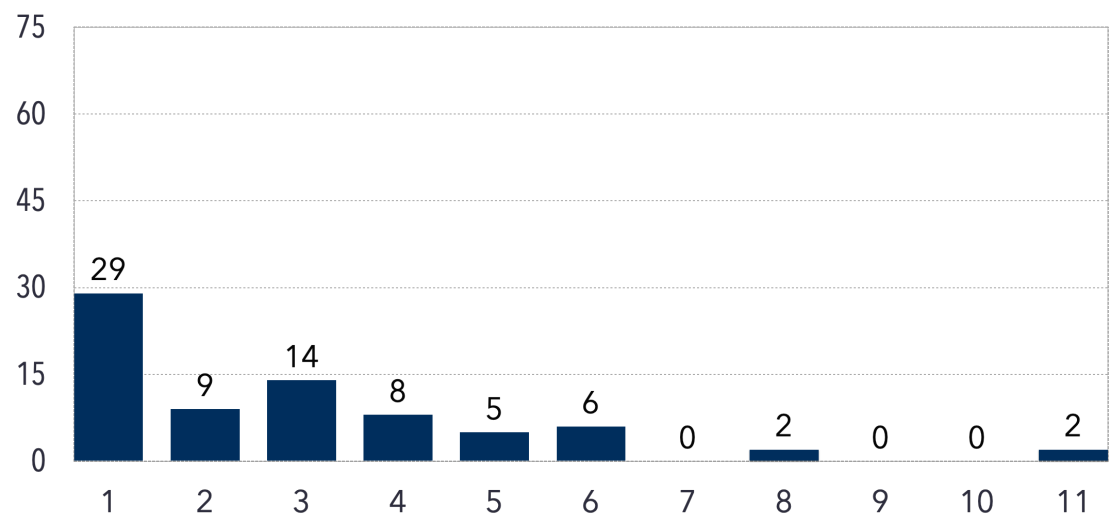 | **2**<br>This statement reflects a sense of denial or a belief in personal invulnerability to the risk of contracting AIDS. It suggests a lack of awareness or concern about the possibility of acquiring the disease. On the scale, it would likely fall towards the negative end, perhaps around 2, indicating a dismissive or unrealistic attitude towards the risk of AIDS. |

## You can get AIDS from toilet seats

| Judges | | | ChatGPT |
|---|---|---|---|
| Median | Interquartile Range | Bar Chart | Category and explanation |
| **2** | **3** | 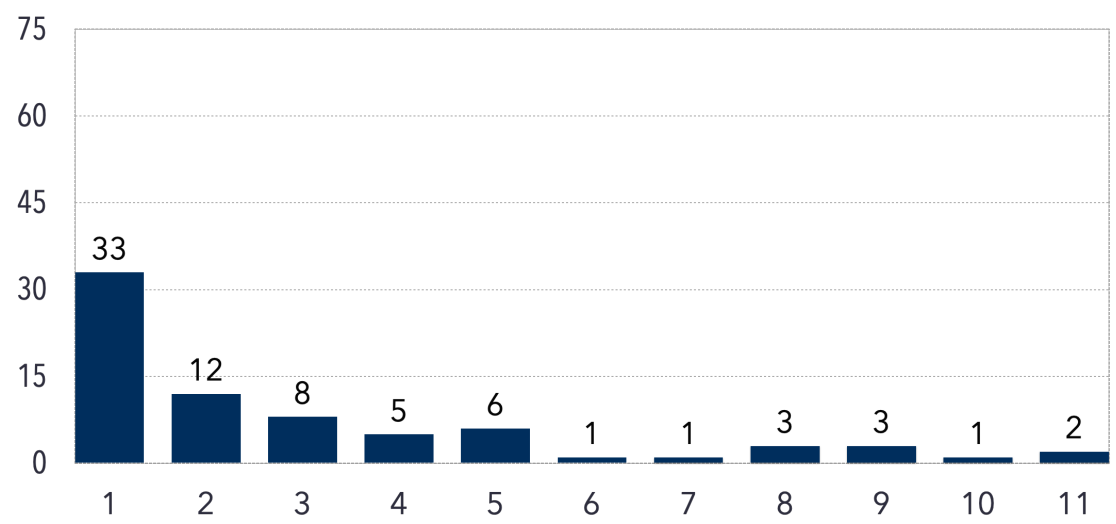 | **1-2**<br>This statement reflects a common misconception about the transmission of HIV/AIDS. In reality, HIV cannot be transmitted through contact with toilet seats. HIV is a fragile virus that cannot survive for long outside of the human body, and it cannot be spread through casual contact with surfaces like toilet seats. Believing that HIV can be contracted in this way perpetuates fear and misinformation about the virus, leading to unnecessary anxiety and stigma. Therefore, this statement would be rated low on the scale, closer to 1, as it represents a false belief that contributes to the misunderstanding of HIV/AIDS transmission. |

**AIDS is a cure, not a disease**

| **Judges** | | | **ChatGPT** |
|---|---|---|---|
| Median | Interquartile Range | Bar Chart | Category and explanation |
| **2** | **4** | 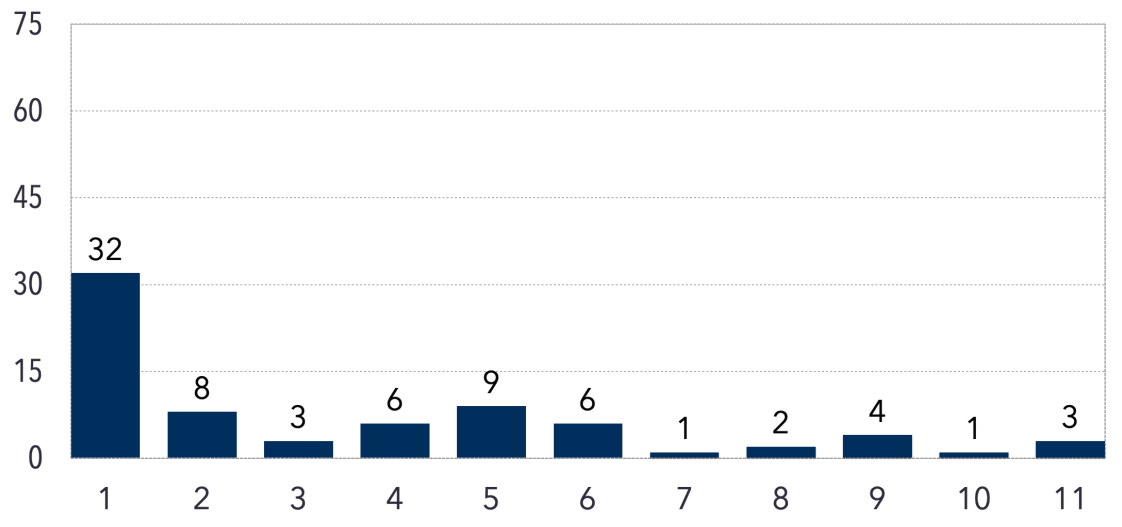 | **1-2**<br>Placing the statement "AIDS is a cure, not a disease" on a scale of 1 to 11, it would fall around 1-2. This rating indicates a highly negative view, as it contradicts established scientific knowledge about AIDS being a serious and life-threatening disease, not a cure. |

**Children cannot catch AIDS**

| **Judges** | | | **ChatGPT** |
|---|---|---|---|
| Median | Interquartile Range | Bar Chart | Category and explanation |
| **2** | **4** | 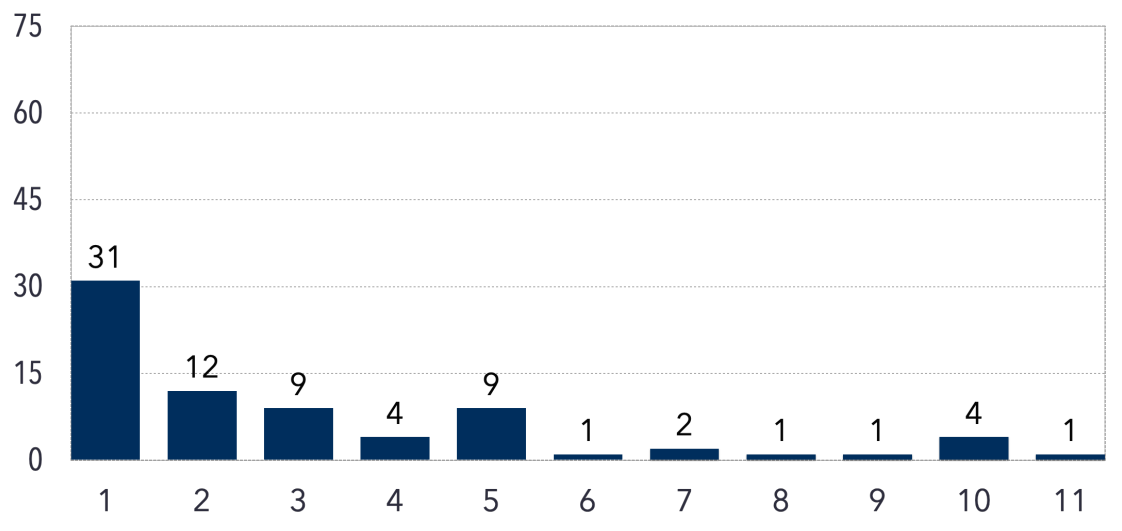 | **2**<br>While the statement might not be as overtly negative or harmful as others, it is still significantly misleading and has potential public health risks. It warrants a rating of 2 on the scale because it perpetuates misinformation, which can have serious consequences for the prevention and treatment of HIV/AIDS in children. |

**Bringing AIDS into my family would be the worst thing I could do**

| **Judges** | | | **ChatGPT** |
|---|---|---|---|
| Median | Interquartile Range | Bar Chart | Category and explanation |
| **2** | **4** | 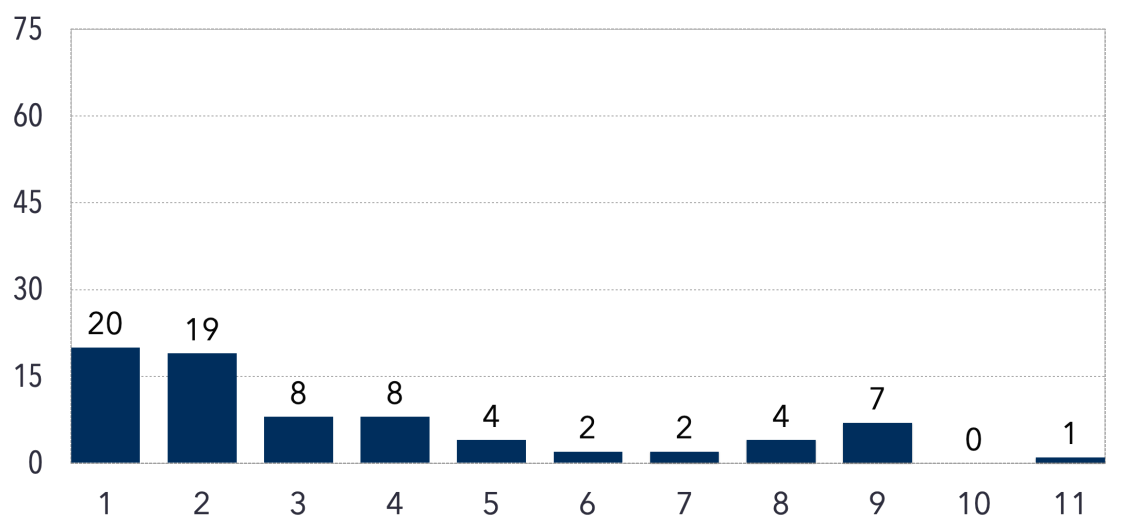 | **2**<br>This statement reflects a fear or concern about the impact of AIDS on one's family and suggests a strong negative sentiment towards the possibility of AIDS affecting them. It implies a perception of AIDS as a highly undesirable and potentially devastating outcome. The statement may evoke feelings of anxiety, worry, or even stigma associated with the disease. Given its strong negative connotation, it would likely be rated around 2 on the scale. |

**The AIDS quilt is an emotional reminder to remember those who did not deserve to die painfully or in vain**

<table>
<tr><th colspan="3">Judges</th><th>ChatGPT</th></tr>
<tr><td>Median</td><td>Interquartile Range</td><td>Bar Chart</td><td>Category and explanation</td></tr>
<tr><td>2</td><td>4</td><td>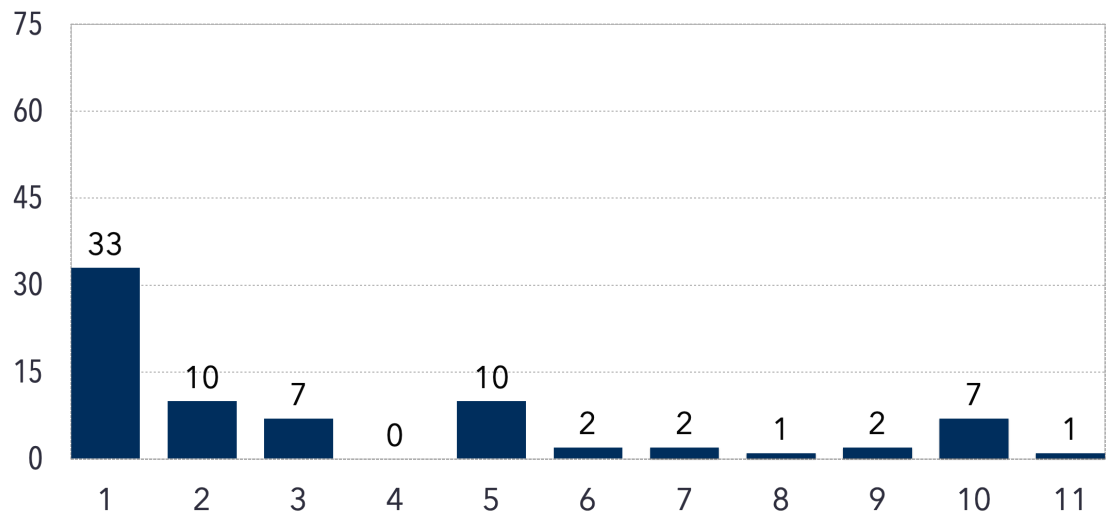
</td><td>10<br>The statement is highly positive as it emphasizes the emotional and educational importance of the AIDS quilt. It recognizes the humanity and dignity of those who died from AIDS, promotes empathy, and underscores the significance of remembering and learning from the epidemic. Therefore, it warrants a rating of 10 on the scale for its compassionate and meaningful acknowledgment of the AIDS quilt's purpose.</td></tr>
</table>

## Category 3

**Very few people have AIDS, so it's unlikely that I'll ever come into contact with a sufferer**

<table>
<tr><th colspan="3">Judges</th><th>ChatGPT</th></tr>
<tr><td>Median</td><td>Interquartile Range</td><td>Bar Chart</td><td>Category and explanation</td></tr>
<tr><td>3</td><td>3</td><td>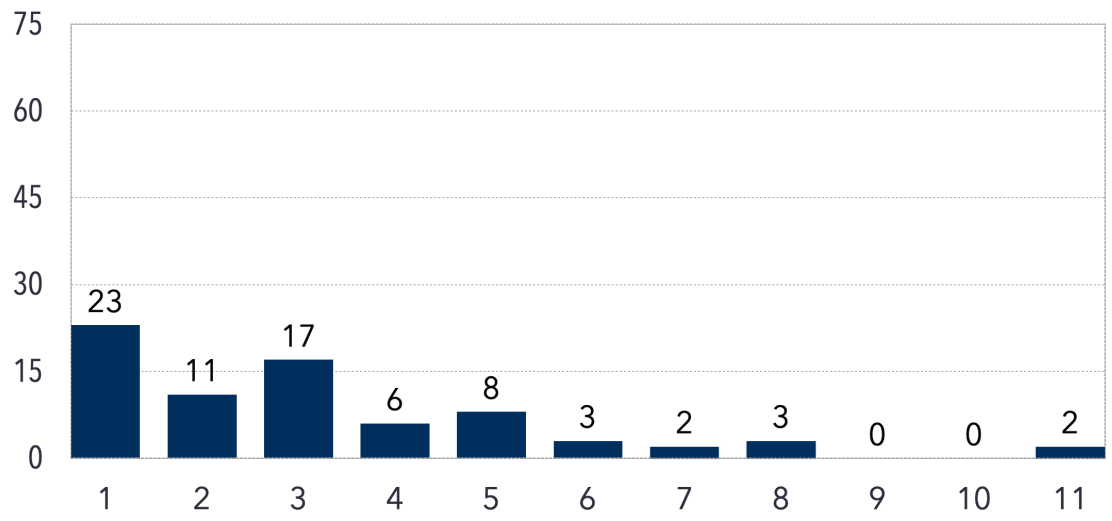
</td><td>3<br>This statement conveys a perception of low personal risk regarding encountering someone with AIDS due to the belief that the prevalence of the disease is rare. It reflects a sense of distance or detachment from the issue, suggesting a lack of concern or awareness about the prevalence of AIDS in society. While not overtly negative towards people with AIDS, it demonstrates a lack of understanding about the widespread impact of the disease. On the scale, it might be rated around 3, indicating a moderately negative attitude due to its disregard for the prevalence and impact of AIDS.</td></tr>
</table>

**By the time I would get sick from AIDS, there will be a cure**

<table>
<tr><th colspan="3">Judges</th><th>ChatGPT</th></tr>
<tr><td>Median</td><td>Interquartile Range</td><td>Bar Chart</td><td>Category and explanation</td></tr>
<tr><td>3</td><td>3</td><td>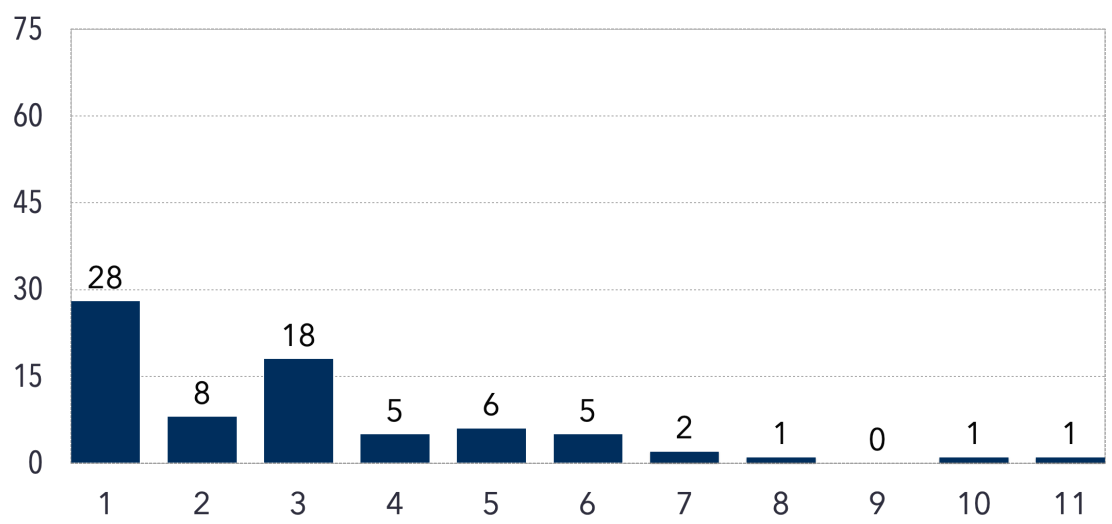
</td><td>3<br>While the statement reflects hope and optimism about future medical advancements, it is misleading and potentially dangerous because it fosters complacency and neglects the importance of prevention and current treatment methods. It warrants a rating of 3 on the scale due to its potential to spread misinformation and undermine effective public health strategies.</td></tr>
</table>

### You can get AIDS from public toilets

| Judges | | | ChatGPT |
|---|---|---|---|
| Median | Interquartile Range | Bar Chart | Category and explanation |
| **3** | **4** | 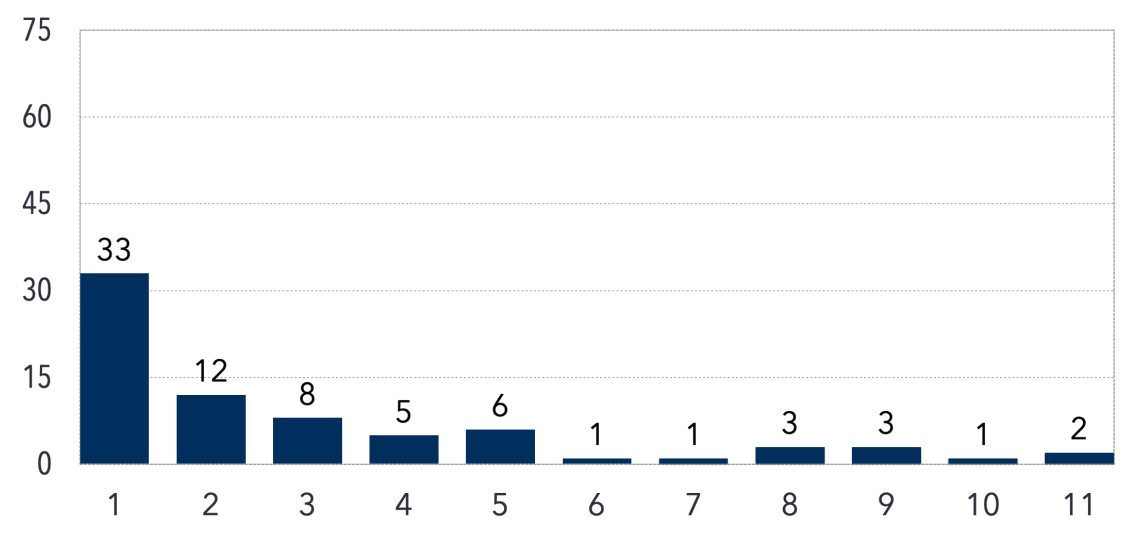 | **3-4**<br>This statement might fall around a 3 or 4 on the scale. It reflects a misconception about the transmission of AIDS, suggesting a lack of understanding or misinformation about how the virus spreads. |

### AIDS is the worst thing that could happen to you

| Judges | | | ChatGPT |
|---|---|---|---|
| Median | Interquartile Range | Bar Chart | Category and explanation |
| **3** | **4** | 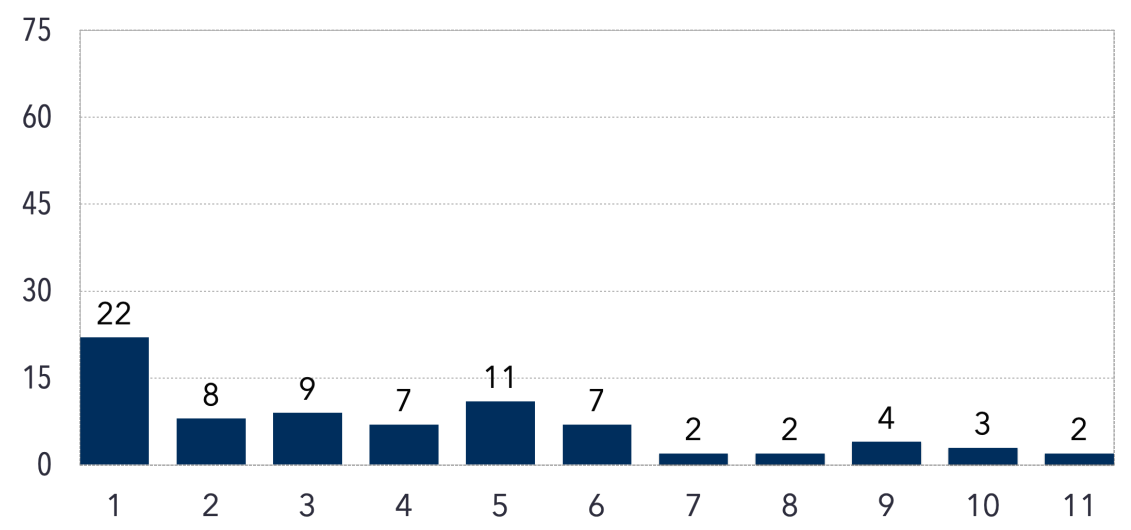 | **2**<br>While recognizing the seriousness of HIV/AIDS, the statement is highly negative and not reflective of the current medical landscape. It exacerbates stigma and fear rather than promoting understanding and support for those affected. Therefore, it warrants a rating of 2 on the scale due to its potential to cause harm through misinformation and stigmatization. |

### You can get AIDS from kissing

| Judges | | | ChatGPT |
|---|---|---|---|
| Median | Interquartile Range | Bar Chart | Category and explanation |
| **3** | **4** | 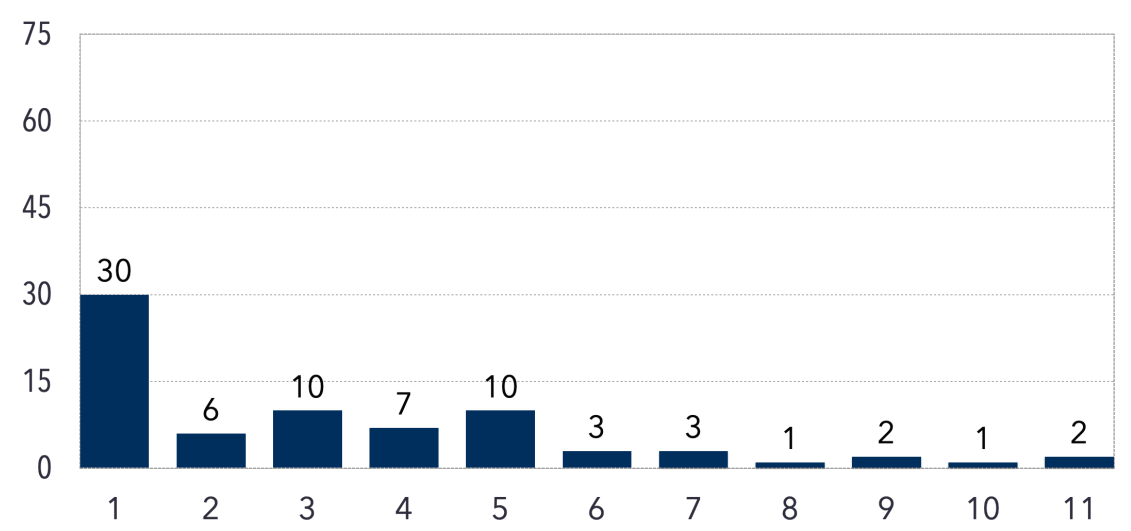 | **2**<br>It perpetuates fear and misinformation about the transmission of HIV. It contributes to stigma and discrimination against people living with HIV/AIDS. Therefore, it would typically be rated low on the scale, around 2. |

### I can't get AIDS if I'm in a monogamous relationship

| Judges | | | ChatGPT |
|---|---|---|---|
| Median | Interquartile Range | Bar Chart | Category and explanation |
| **3** | **4** | 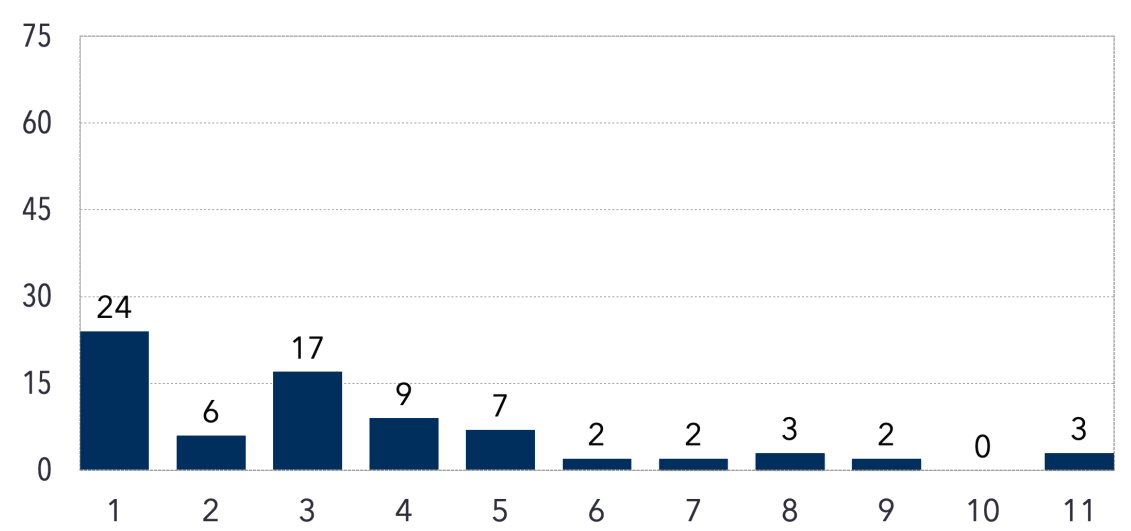 | **3**<br>While monogamy can reduce the risk of HIV transmission within a relationship, it's important to recognize its limitations and promote comprehensive prevention strategies. By emphasizing the importance of monogamy within the context of broader HIV prevention efforts, we can encourage informed decision-making and reduce stigma. Therefore, I'll rate it a 3. |

**The AIDS virus will not ever affect me**

| **Judges** | | | **ChatGPT** |
|---|---|---|---|
| Median | Interquartile Range | Bar Chart | Category and explanation |
| **3** | **4** | 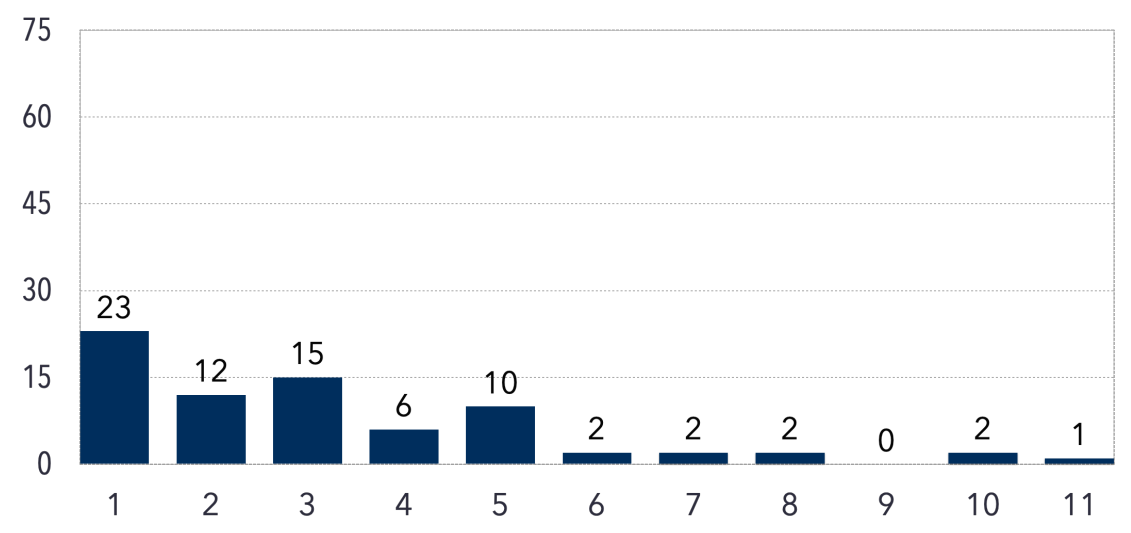 | **2**<br>This statement reflects a sense of personal invulnerability to the AIDS virus, suggesting a lack of concern or awareness about the risks associated with it. Such a perspective could be considered somewhat naive or dismissive of the potential impact of the virus. Therefore, it would be rated negatively, perhaps around 2, for its lack of acknowledgment of the potential risks and impact of AIDS. |

**AIDS is becoming more a problem for heterosexual women and their offsprings than IV drug users or homosexuals**

| **Judges** | | | **ChatGPT** |
|---|---|---|---|
| Median | Interquartile Range | Bar Chart | Category and explanation |
| **4** | **3** | 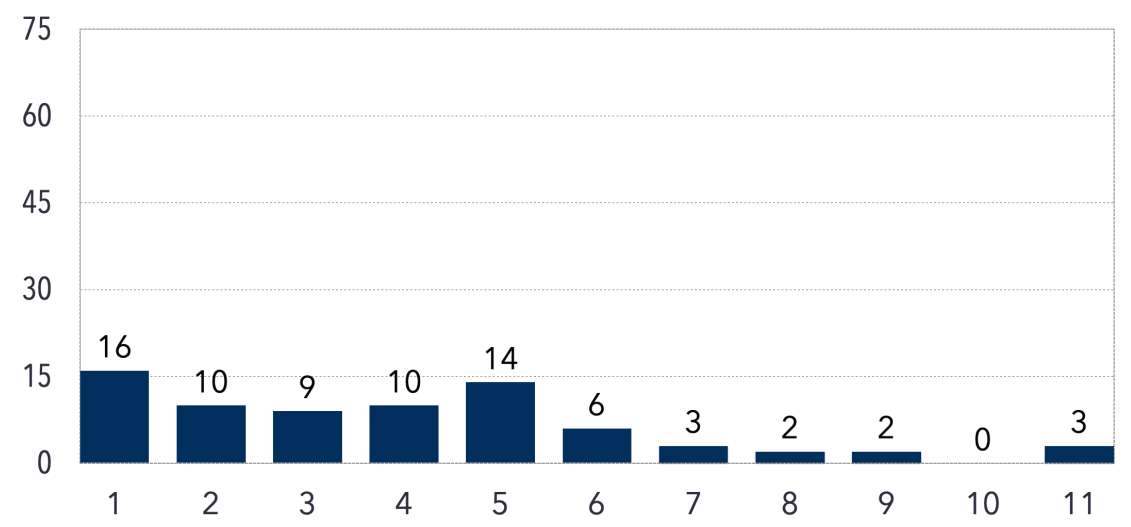 | **3**<br>While acknowledging the importance of addressing HIV/AIDS among heterosexual women and their offspring, the statement oversimplifies the issue and may contribute to stigmatization and misunderstanding of HIV/AIDS epidemiology. It's crucial to approach HIV/AIDS prevention and treatment with nuance, recognizing the diverse experiences and vulnerabilities of different populations. Therefore, I'll rate it a 3 on the scale. |

**You can't get AIDS from oral sex**

| **Judges** | | | **ChatGPT** |
|---|---|---|---|
| Median | Interquartile Range | Bar Chart | Category and explanation |
| **4** | **4** | 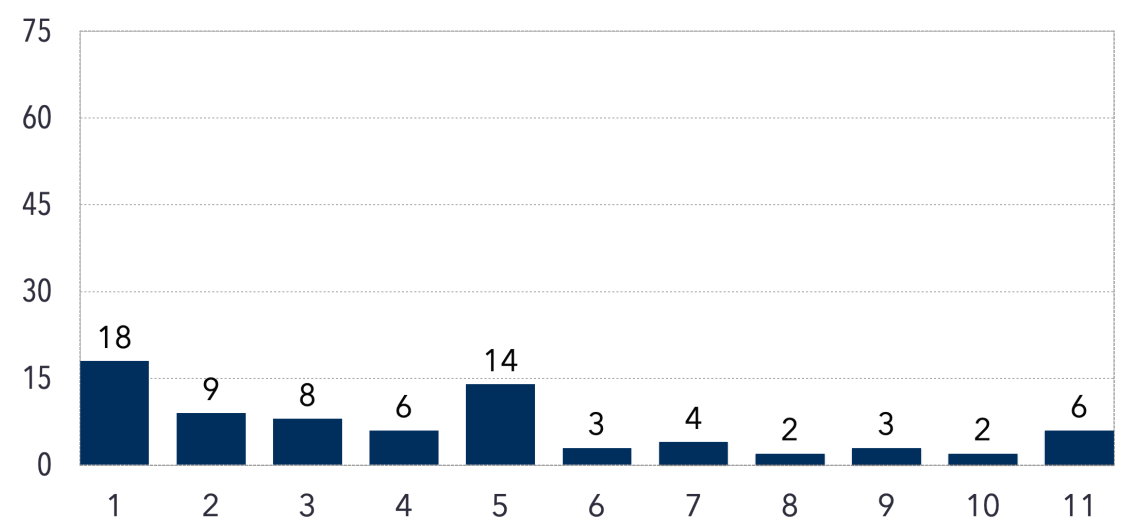 | **5**<br>This statement is a misconception. While the risk of HIV transmission through oral sex is generally lower compared to vaginal or anal sex, it is still possible. HIV can be transmitted through oral sex if there are open sores or cuts in the mouth, or if there is ejaculation into the mouth. Therefore, it's important to use protection such as condoms or dental dams to reduce the risk. On the scale, this statement might fall around 5, indicating some awareness of the reduced risk but also some misunderstanding about the potential transmission. |

**Because AIDS is preventable, we should focus our resources on prevention instead of curing**

| **Judges** | | | **ChatGPT** |
|---|---|---|---|
| Median | Interquartile Range | Bar Chart | Category and explanation |
| **4** | **5** | 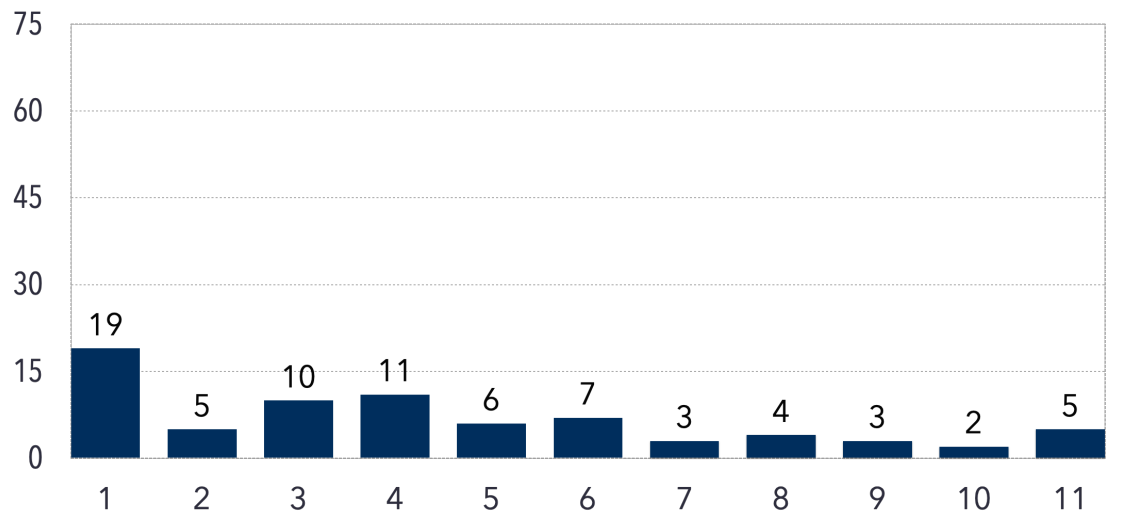 | **8**<br>This statement could be rated around 8. It suggests prioritizing prevention efforts over finding a cure for AIDS. While prevention is crucial in reducing the spread of HIV/AIDS, it's also essential to continue investing resources in research for a cure or better treatments. Preventive measures alone may not eliminate HIV/AIDS entirely, and advancements in treatment can significantly improve the quality of life for those living with the disease." |

**People with AIDS deserve our understanding, but not necessarily special treatment**

| **Judges** | | | **ChatGPT** |
|---|---|---|---|
| Median | Interquartile Range | Bar Chart | Category and explanation |
| **4** | **5** | 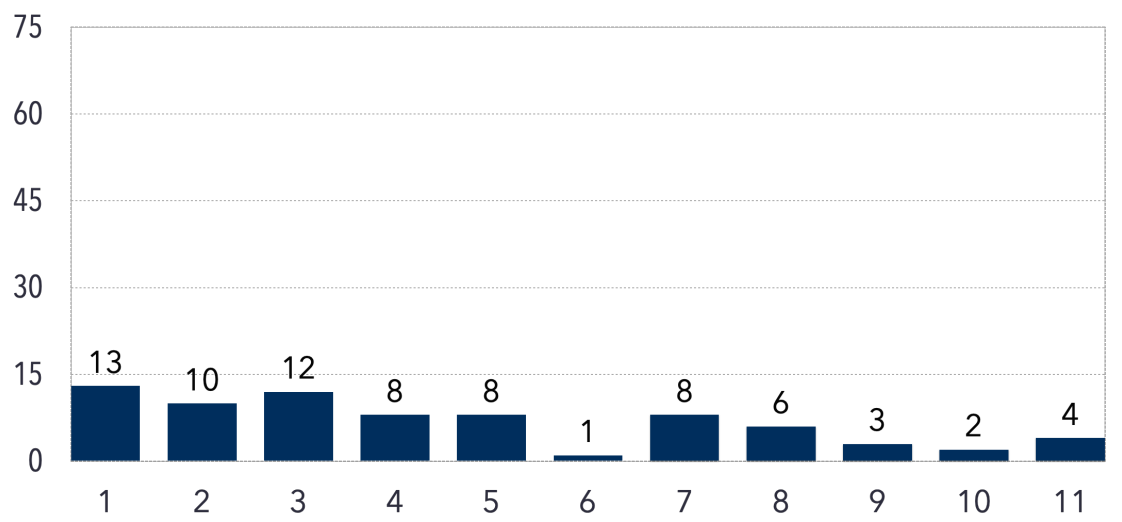 | **5-6**<br>This statement suggests a somewhat neutral or ambivalent attitude towards individuals with AIDS. While it acknowledges the importance of understanding, it also implies a reluctance to provide special treatment or accommodations for those with the disease. On the scale, it might be rated around 5 or 6, indicating a middling stance between positive and negative perceptions of people with AIDS. |

## Category 5

**Universal precautions are infallible**

| **Judges** | | | **ChatGPT** |
|---|---|---|---|
| Median | Interquartile Range | Bar Chart | Category and explanation |
| **5** | **3** | 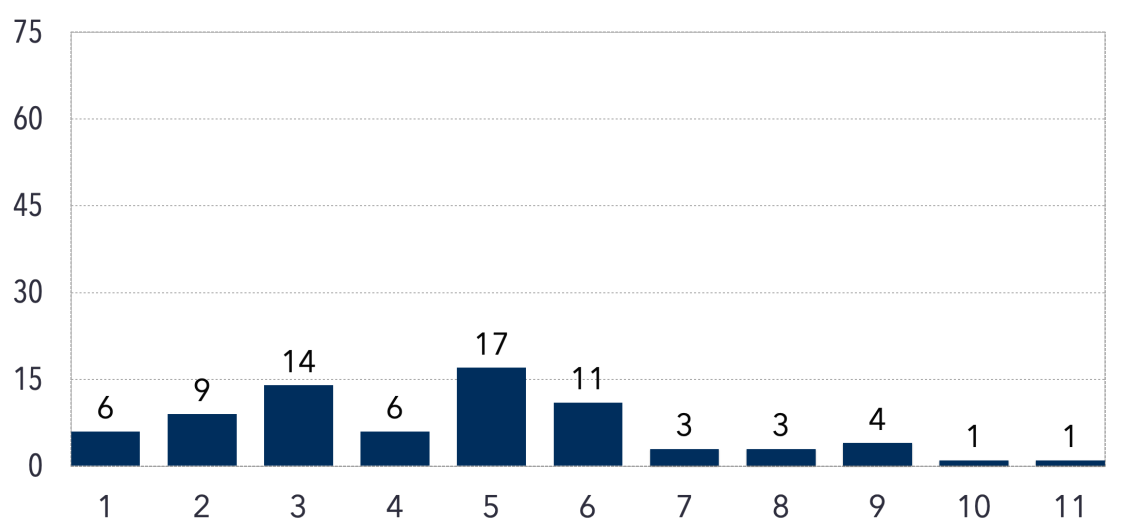 | **4**<br>While universal precautions are highly effective and crucial for infection control, they are not infallible. Acknowledging their high efficacy while recognizing the potential for human error and the need for complementary measures provides a more balanced view. Therefore, the statement warrants a rating of 4 on the scale. |

**People infected with AIDS did not have safe sex**

| **Judges** | | | **ChatGPT** |
|---|---|---|---|
| Median | Interquartile Range | Bar Chart | Category and explanation |
| **5** | **3** | 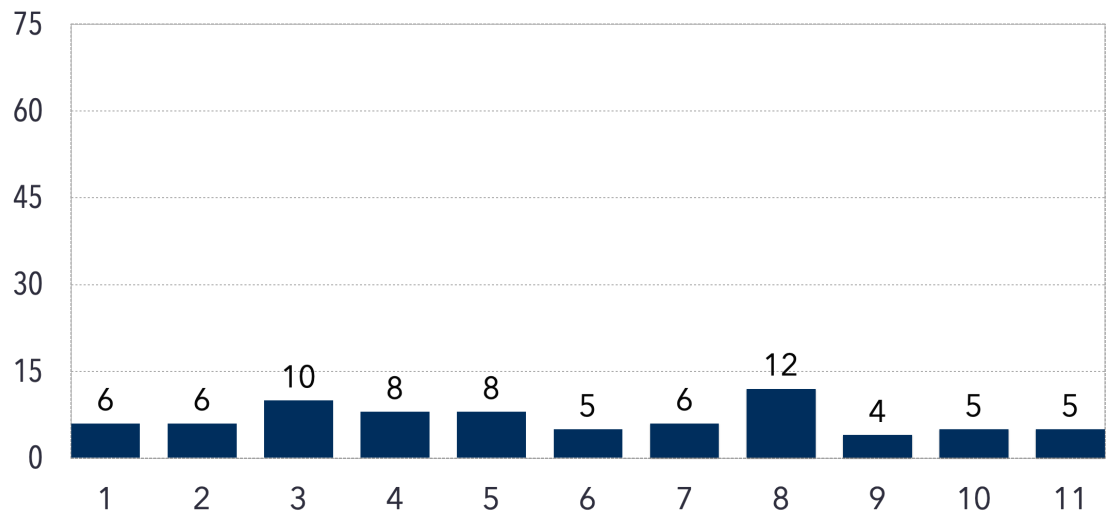 | **2-3**<br>This statement carries a negative connotation as it implies blame or judgment towards individuals with AIDS for their infection. It suggests a lack of empathy or understanding of the complex factors surrounding HIV transmission. Therefore, it would be rated negatively, perhaps around 2 or 3, for its stigmatizing attitude towards people with AIDS. |

**The likelihood of contracting AIDS is very low**

| **Judges** | | | **ChatGPT** |
|---|---|---|---|
| Median | Interquartile Range | Bar Chart | Category and explanation |
| **5** | **3** | 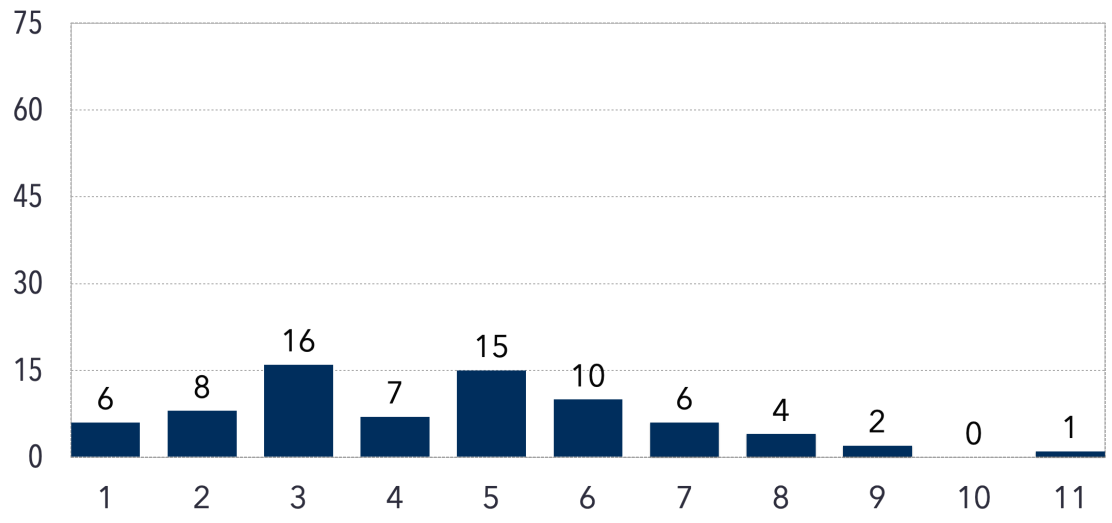 | **5-6**<br>This statement implies a relatively positive perception of the risk of contracting AIDS, suggesting that it is not very common. It may lead to a sense of complacency or underestimation of the importance of preventive measures. Therefore, it would be rated moderately negative, perhaps around 5 or 6, for downplaying the importance of AIDS prevention and awareness efforts. |

**The number of individuals with AIDS in Hollywood is higher than the general public thinks**

| **Judges** | | | **ChatGPT** |
|---|---|---|---|
| Median | Interquartile Range | Bar Chart | Category and explanation |
| **5** | **3** | 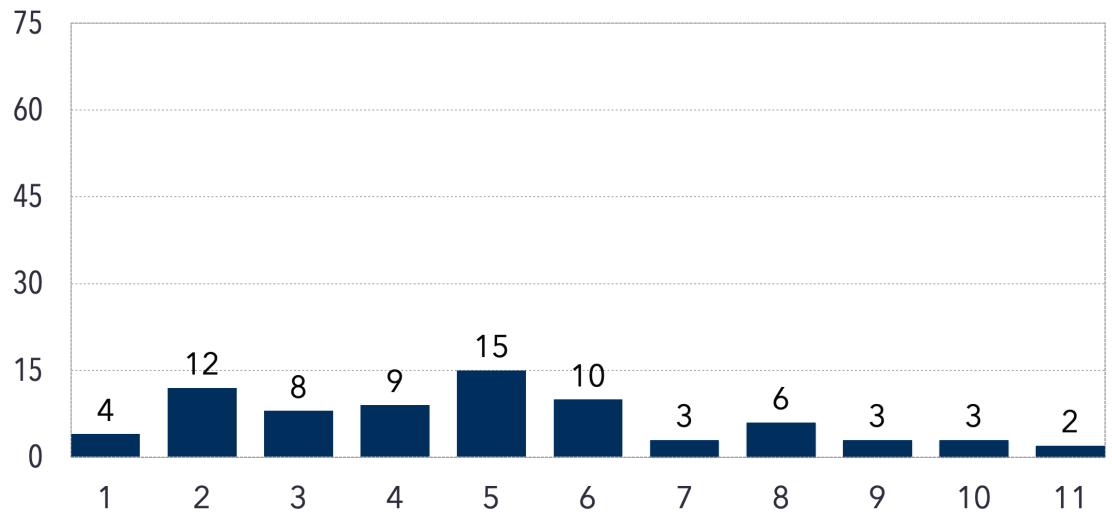 | **5**<br>This statement suggests that there may be a higher prevalence of individuals with AIDS in Hollywood than what is commonly believed. It implies a sense of awareness and recognition of the issue within a specific community. The rating for this statement would likely fall in the middle range, around 5 or 6, as it acknowledges a concerning situation but does not necessarily provide a positive or negative judgment. |

## Category 6

### It's easy to get AIDS

| **Judges** | | | **ChatGPT** |
|---|---|---|---|
| Median | Interquartile Range | Bar Chart | Category and explanation |
| **6** | **3** | 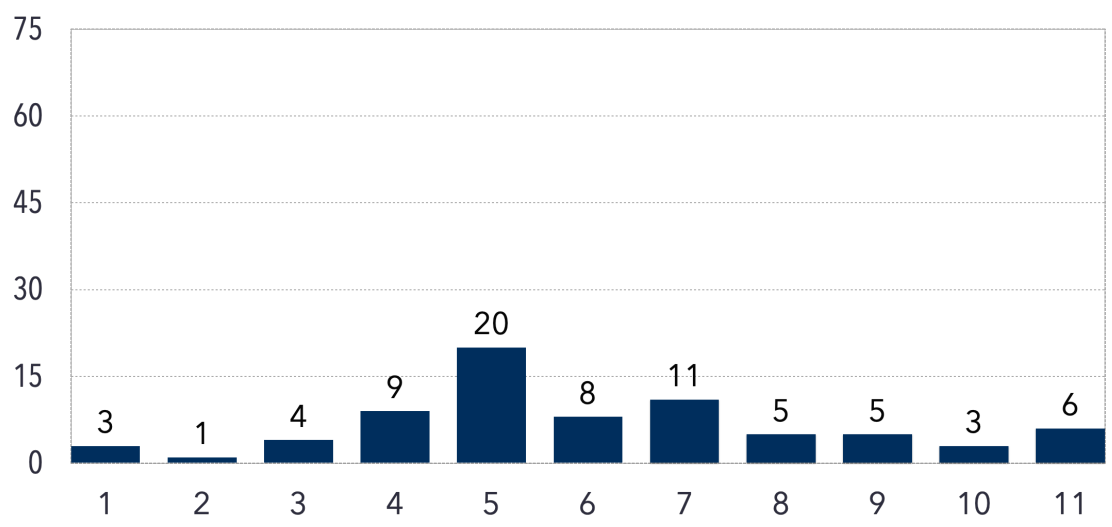  | **3** This statement suggests a negative perception of AIDS, indicating that it is easily contracted. It may contribute to fear and misunderstanding about the transmission of the disease. Therefore, it would be rated negatively, perhaps around 3, for potentially perpetuating misconceptions and stigma surrounding AIDS. |

## Category 7

### A cure for AIDS is on the horizon

| **Judges** | | | **ChatGPT** |
|---|---|---|---|
| Median | Interquartile Range | Bar Chart | Category and explanation |
| **7** | **4** | 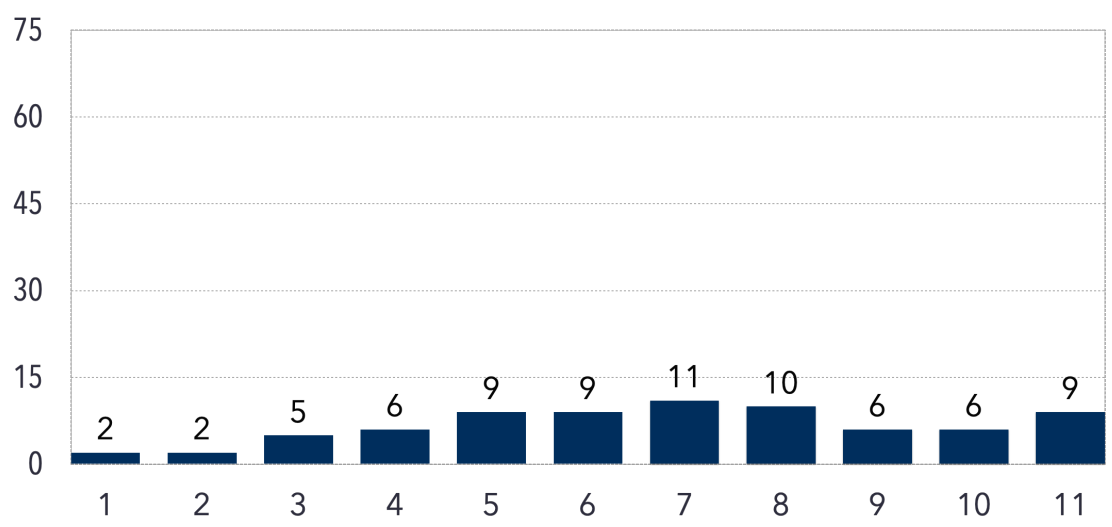  | **7** The statement reflects a positive and hopeful outlook based on significant scientific advancements and ongoing research. It serves to inspire confidence and motivate continued efforts while recognizing the importance of balancing optimism with realistic expectations. Therefore, the statement warrants a rating of 7 on the scale for its constructive and forward-looking perspective. |

### AIDS is an epidemic that affects us all

| **Judges** | | | **ChatGPT** |
|---|---|---|---|
| Median | Interquartile Range | Bar Chart | Category and explanation |
| **7** | **6** | 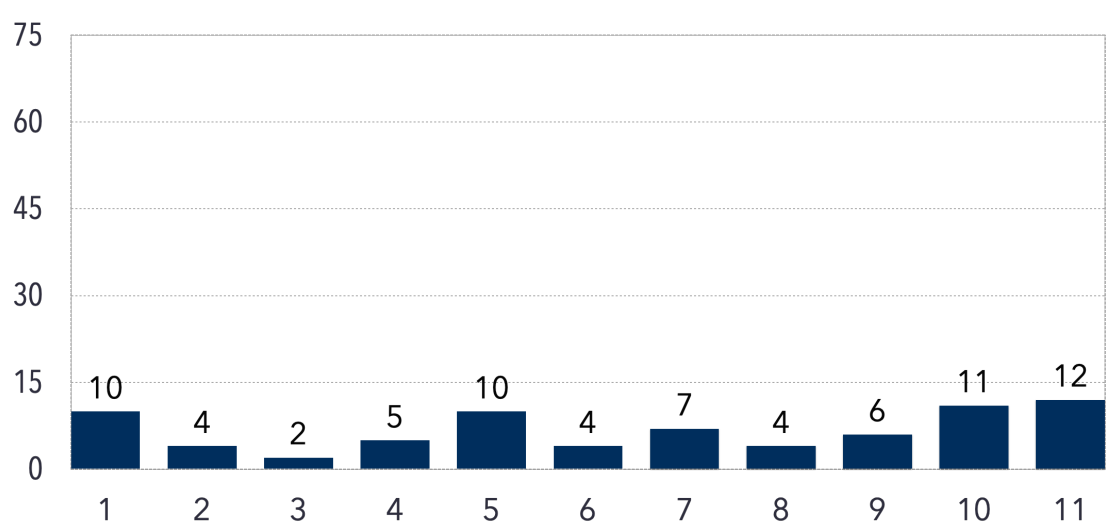  | **8** The statement is positive and important as it recognizes the global impact of HIV/AIDS, promotes collective responsibility and solidarity, and emphasizes the need for widespread awareness and education. Therefore, it warrants a rating of 8 on the scale for its constructive implications and the call to action it represents. |

## Category 8

### If you have AIDS, you can still live a normal life

| **Judges** | | | **ChatGPT** |
|---|---|---|---|
| Median | Interquartile Range | Bar Chart | Category and explanation |
| **8** | **4** | 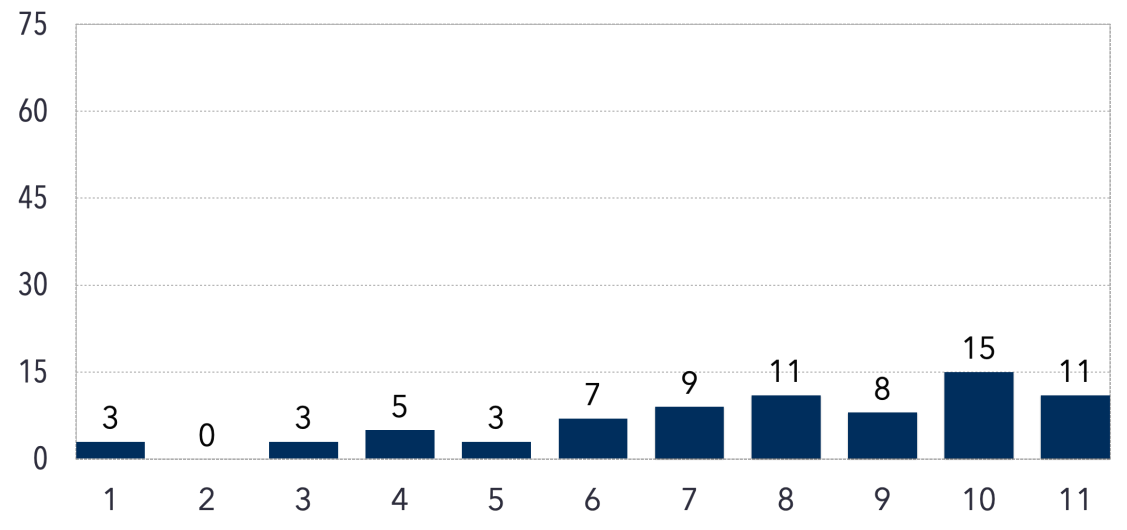 | **7-8**<br>This statement suggests a more positive and hopeful outlook regarding living with AIDS, indicating that it's possible to lead a normal life despite the diagnosis. Given its relatively optimistic tone compared to other statements, it could be rated around 7-8 on the scale. |

### People with AIDS are like my parents

| **Judges** | | | **ChatGPT** |
|---|---|---|---|
| Median | Interquartile Range | Bar Chart | Category and explanation |
| **8** | **5** | 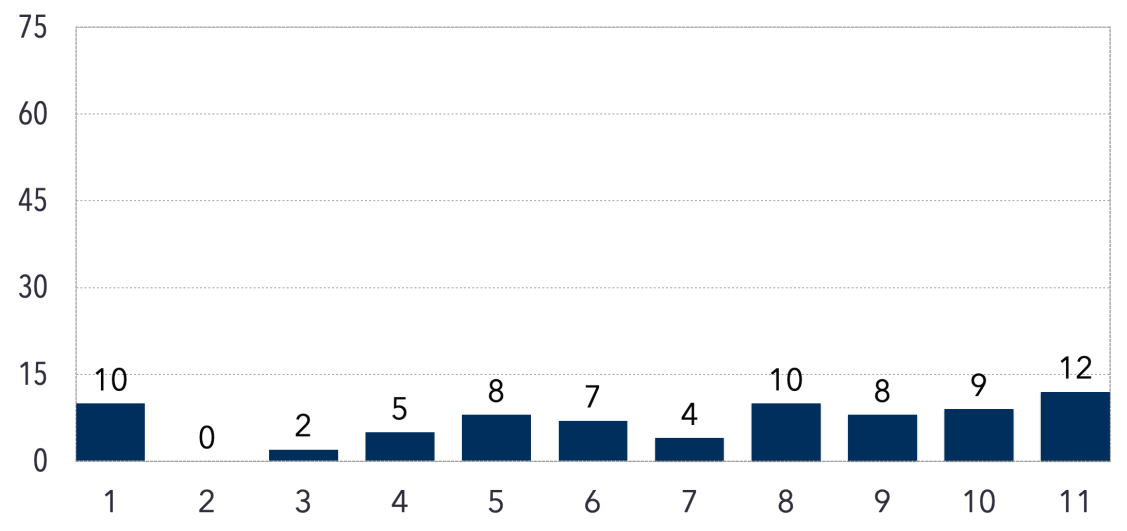 | **7-8**<br>It reflects a relatively positive perception, suggesting empathy and understanding towards individuals with AIDS by likening them to someone close, such as parents. However, it might also carry some weight of personal bias, potentially limiting broader understanding and empathy for all people affected by AIDS. |

### The nation's blood supply is safe

| **Judges** | | | **ChatGPT** |
|---|---|---|---|
| Median | Interquartile Range | Bar Chart | Category and explanation |
| **8** | **5** | 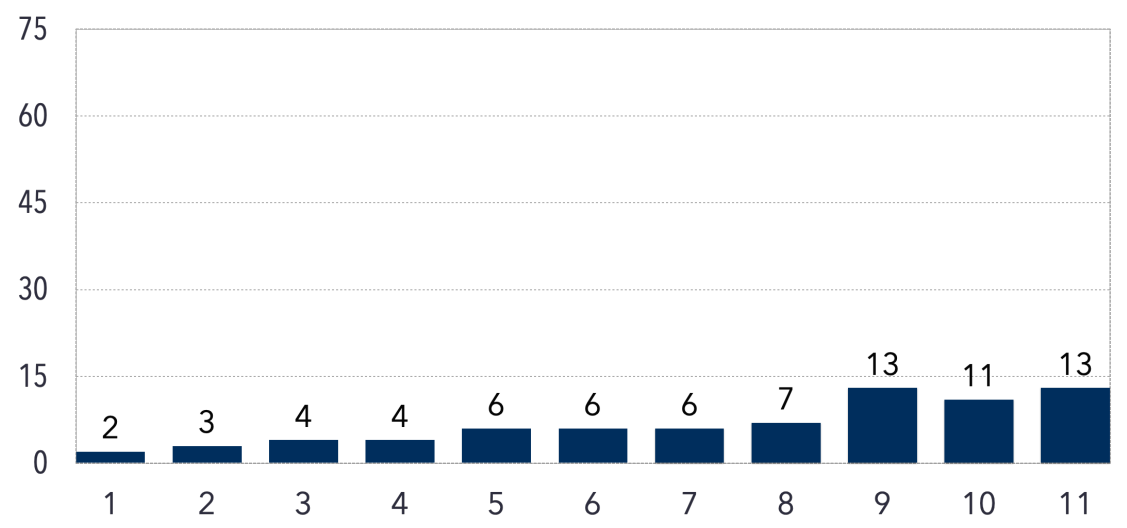 | **7-8**<br>This statement implies confidence in the safety of the nation's blood supply regarding the transmission of AIDS. Historically, there have been concerns about the safety of blood transfusions in relation to AIDS transmission, but significant advancements in screening and testing have greatly reduced this risk. On the scale, it would likely be rated moderately. |

**It is not the AIDS virus that kills people, it is complications from other illnesses (because the immune system isn't functioning) that cause death**

| **Judges** | | | **ChatGPT** |
|---|---|---|---|
| Median | Interquartile Range | Bar Chart | Category and explanation |
| **8** | **5** | 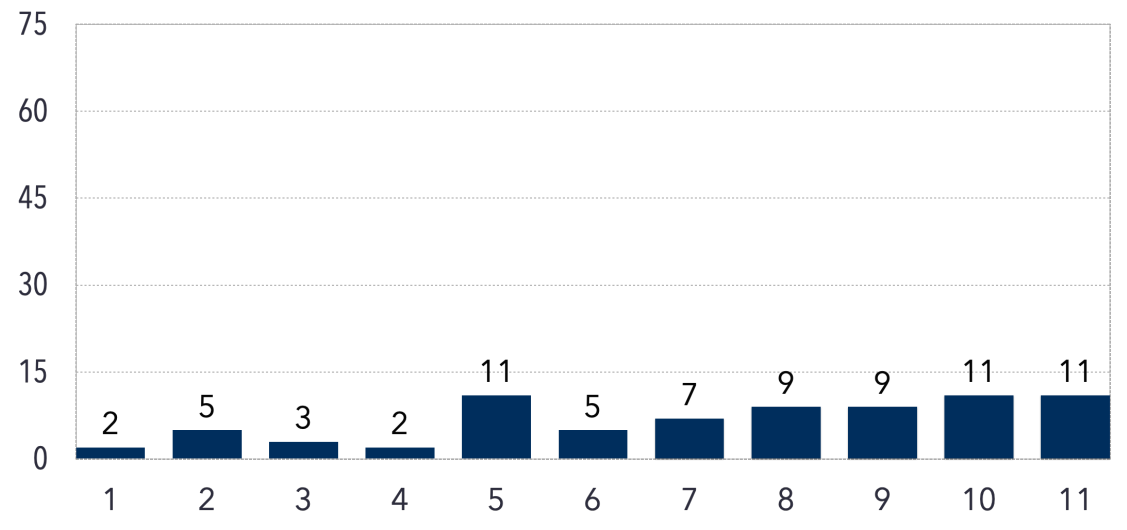 | **8**<br>The statement is both accurate and educational. It helps to clarify the nature of AIDS and its impact on the immune system, promoting a better understanding and encouraging comprehensive care and preventive measures. Therefore, it warrants a rating of 8 on the scale for its informative and constructive value. |

**I know enough about the spread of the disease that I would have no problem working in a health care setting with patients with AIDS**

| **Judges** | | | **ChatGPT** |
|---|---|---|---|
| Median | Interquartile Range | Bar Chart | Category and explanation |
| **8** | **6** | 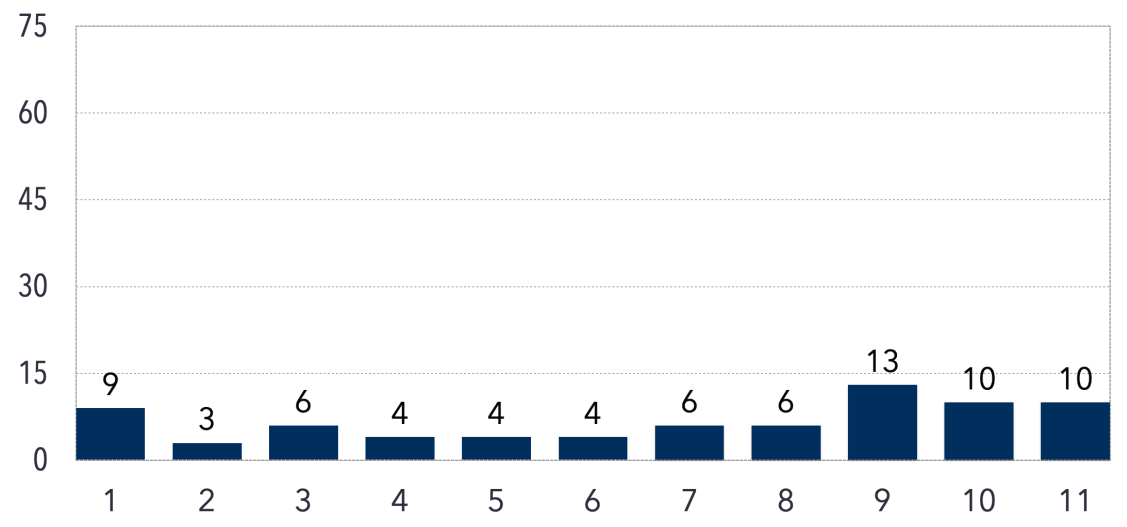 | **8**<br>This statement suggests confidence in one's knowledge about AIDS transmission and a willingness to work with patients with AIDS, indicating a positive attitude towards understanding and acceptance. Therefore, it would be rated positively, around 8, for showing compassion and a lack of stigma towards individuals with AIDS. |

**Mandatory HIV testing should be established for all pregnant women**

| **Judges** | | | **ChatGPT** |
|---|---|---|---|
| Median | Interquartile Range | Bar Chart | Category and explanation |
| **8** | **6** | 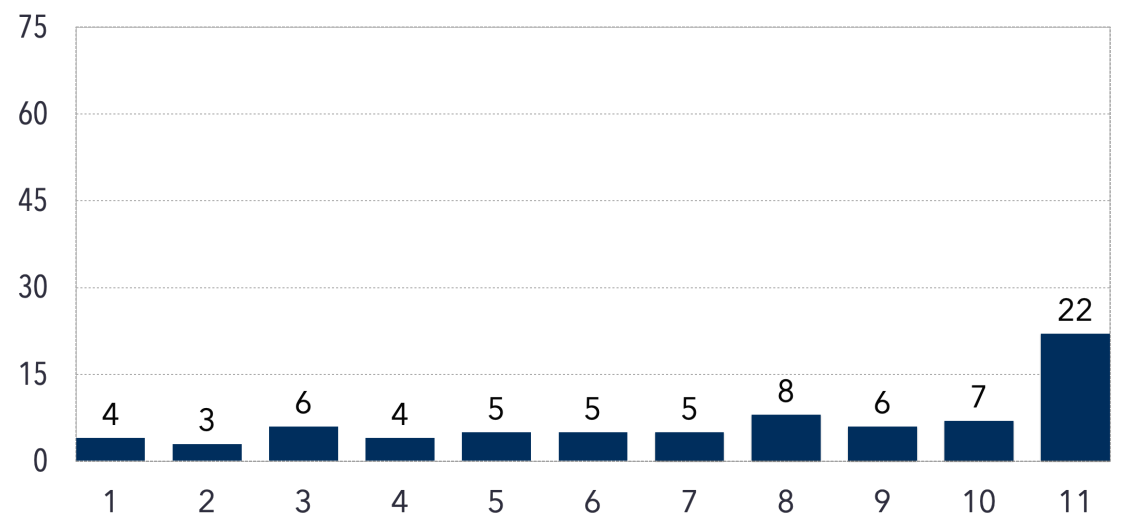 | **7**<br>This statement suggests implementing mandatory HIV testing for all pregnant women. It reflects a belief that such testing could have positive outcomes, such as preventing mother-to-child transmission of HIV. On the rating scale, it might be considered as around 7, indicating a proactive approach to addressing HIV/AIDS but also raising questions about personal autonomy and medical ethics. |

### Category 9

**Condoms will always prevent the spread of AIDS**

| **Judges** | | | **ChatGPT** |
|---|---|---|---|
| Median | Interquartile Range | Bar Chart | Category and explanation |
| **9** | **5** | 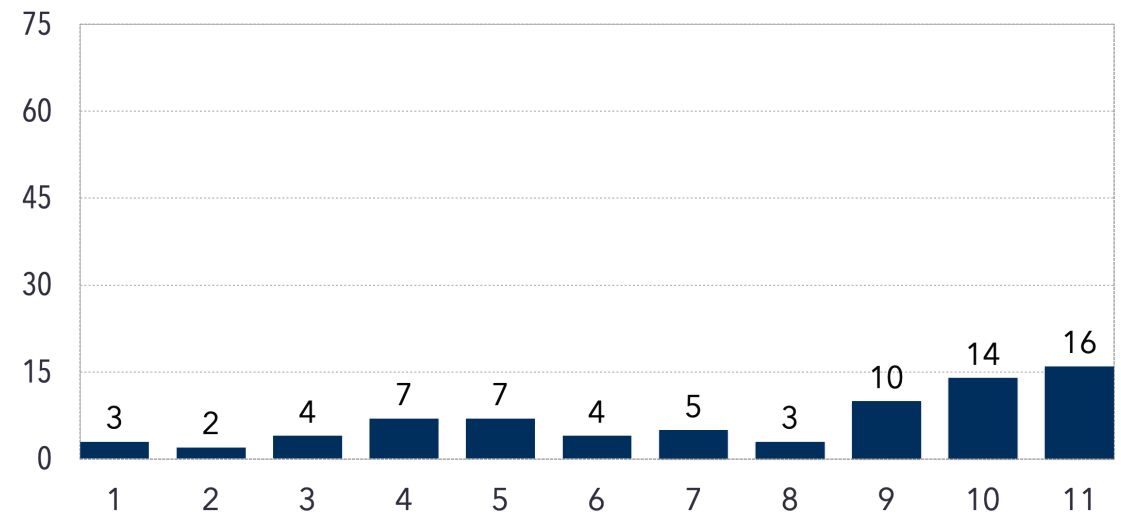 | **8**<br>Generally, it could be categorized around 8, reflecting a positive but not absolute assurance given the common understanding that condoms significantly reduce, but do not entirely eliminate, the risk of spreading AIDS. |

**Aids affects us all**

| **Judges** | | | **ChatGPT** |
|---|---|---|---|
| Median | Interquartile Range | Bar Chart | Category and explanation |
| **9** | **5** | 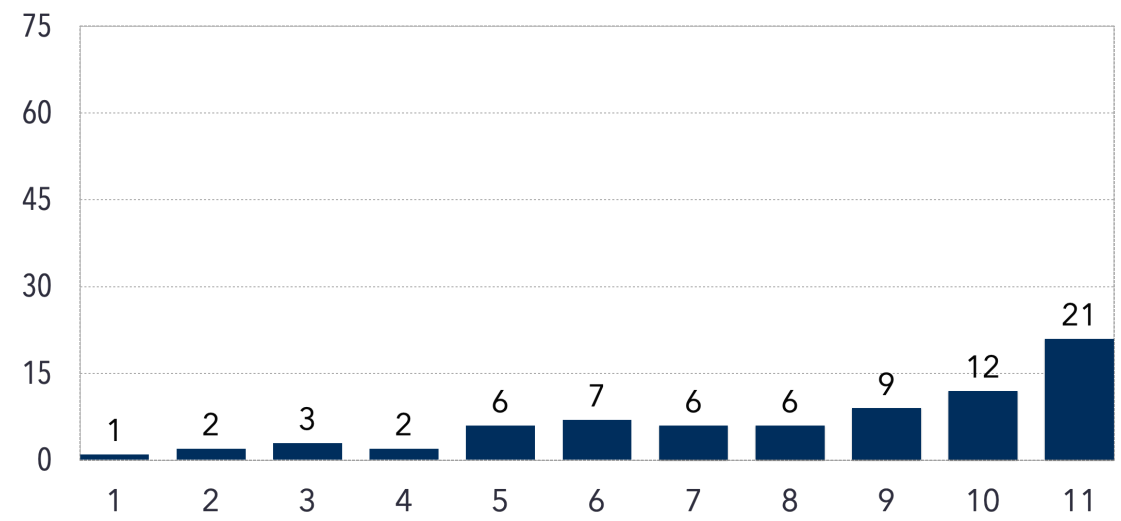 | **9-10**<br>This statement reflects a more inclusive and empathetic view towards AIDS, recognizing its impact on society as a whole. It acknowledges that AIDS is not limited to certain groups or individuals but can affect anyone. Therefore, it would be rated positively, around 9 or 10, for its acknowledgment of the widespread impact of AIDS. |

### Category 10

**People with AIDS should be treated just like everybody else**

| **Judges** | | | **ChatGPT** |
|---|---|---|---|
| Median | Interquartile Range | Bar Chart | Category and explanation |
| **10** | **2** | 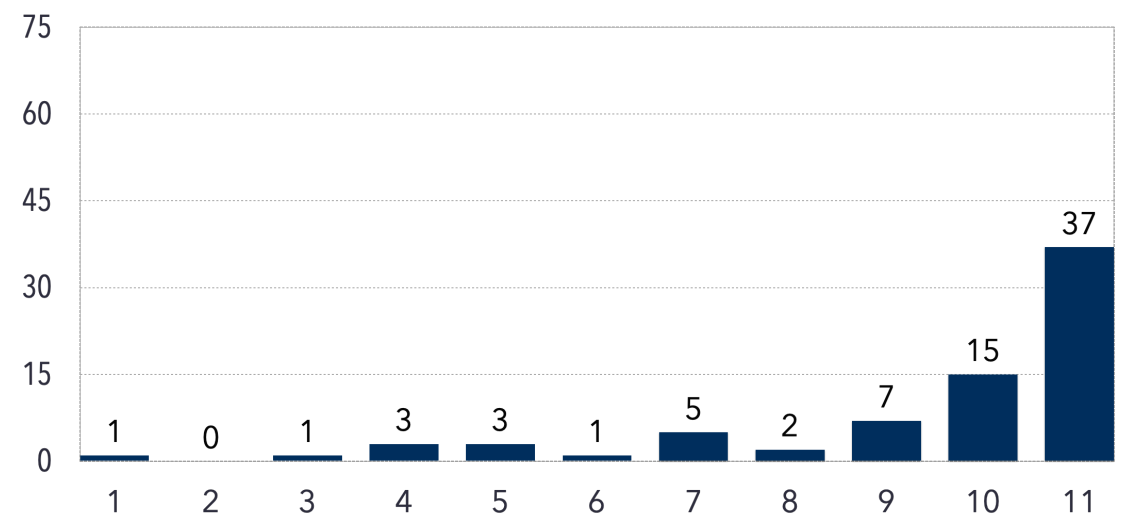 | **9-10**<br>This statement emphasizes equality and inclusivity, advocating for fair treatment of individuals with AIDS. It suggests that people with AIDS should not face discrimination or stigmatization but should receive the same respect and opportunities as everyone else. Therefore, it would be rated positively, around 9 or 10, for promoting compassion and equality. |

**Someone with AIDS could be just like me**

| Judges | | | ChatGPT |
|---|---|---|---|
| Median | Interquartile Range | Bar Chart | Category and explanation |
| **10** | **2** | 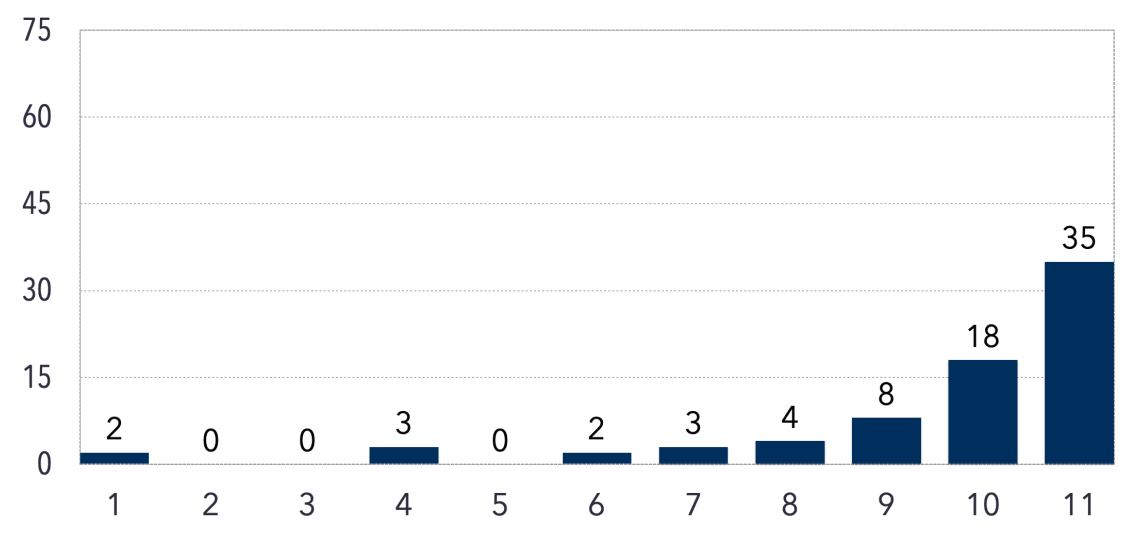 | **10**<br>This statement implies empathy and recognition of the shared humanity with individuals affected by AIDS. It suggests an understanding that anyone, regardless of their background or circumstances, could be impacted by the disease. Therefore, it would be rated positively, around 10, for its acknowledgment of commonality and empathy towards those affected by AIDS. |

**People with AIDS are real people**

| Judges | | | ChatGPT |
|---|---|---|---|
| Median | Interquartile Range | Bar Chart | Category and explanation |
| **10** | **3** | 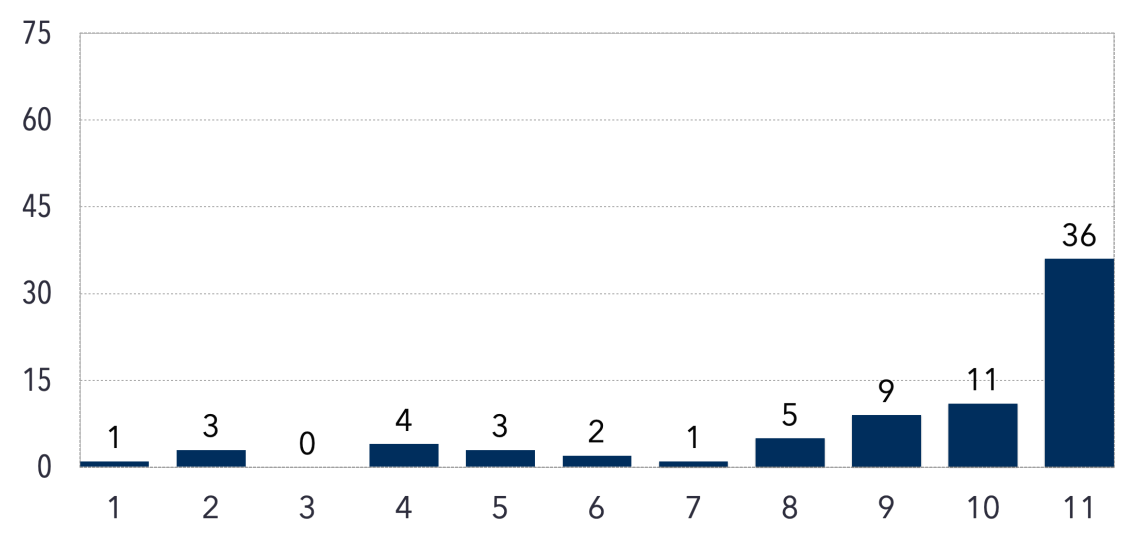 | **9-10**<br>It would fall around 9-10. This rating suggests a highly positive view, recognizing the fundamental humanity and worth of individuals living with AIDS. |

**You can get AIDS from heterosexual sex**

| Judges | | | ChatGPT |
|---|---|---|---|
| Median | Interquartile Range | Bar Chart | Category and explanation |
| **10** | **3** | 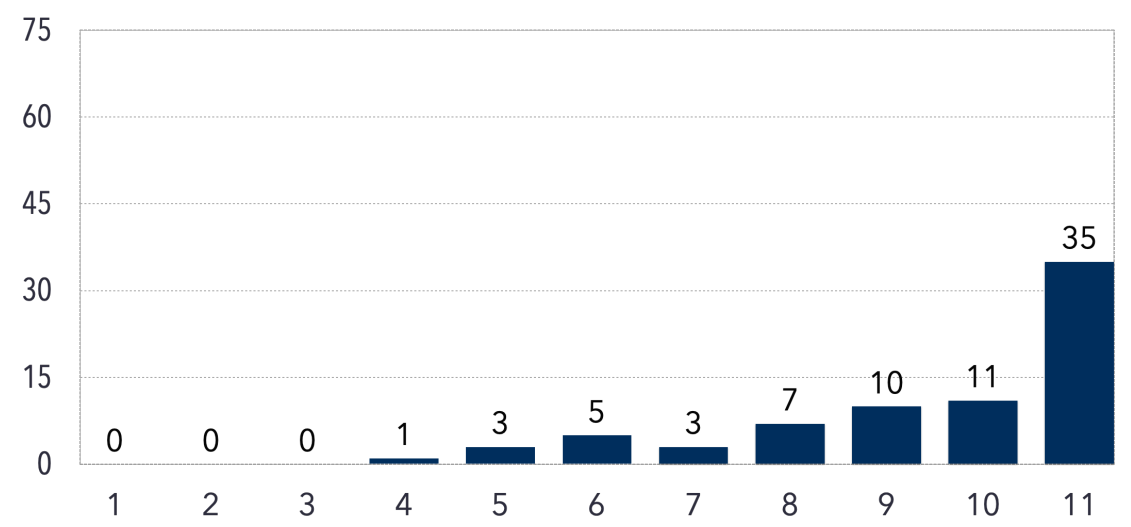 | **6**<br>A rating closer to 6 on the scale would be appropriate. This placement acknowledges that the statement contains some accurate information but may also perpetuate misconceptions or oversimplify the issue. |

**AIDS does not discriminate**

| Judges | | | ChatGPT |
|---|---|---|---|
| Median | Interquartile Range | Bar Chart | Category and explanation |
| **10** | **3** | 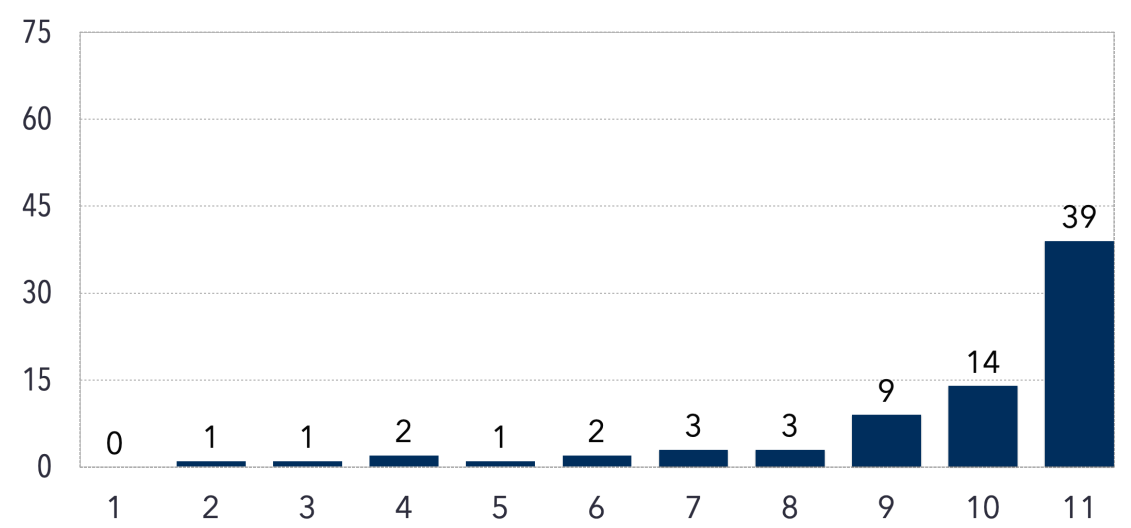 | **10**<br>AIDS does not discriminate based on factors such as race, gender, sexual orientation, or socioeconomic status. However, certain populations may be disproportionately affected by HIV/AIDS due to social determinants of health, including access to healthcare, education, and prevention services. Overall, this statement would rate high on the scale, around 10, for its recognition of the non-discriminatory nature of AIDS. |

**AIDS is a disease that anyone can get if there are not careful**

| **Judges** | | | **ChatGPT** |
|---|---|---|---|
| Median | Interquartile Range | Bar Chart | Category and explanation |
| **10** | **5** | 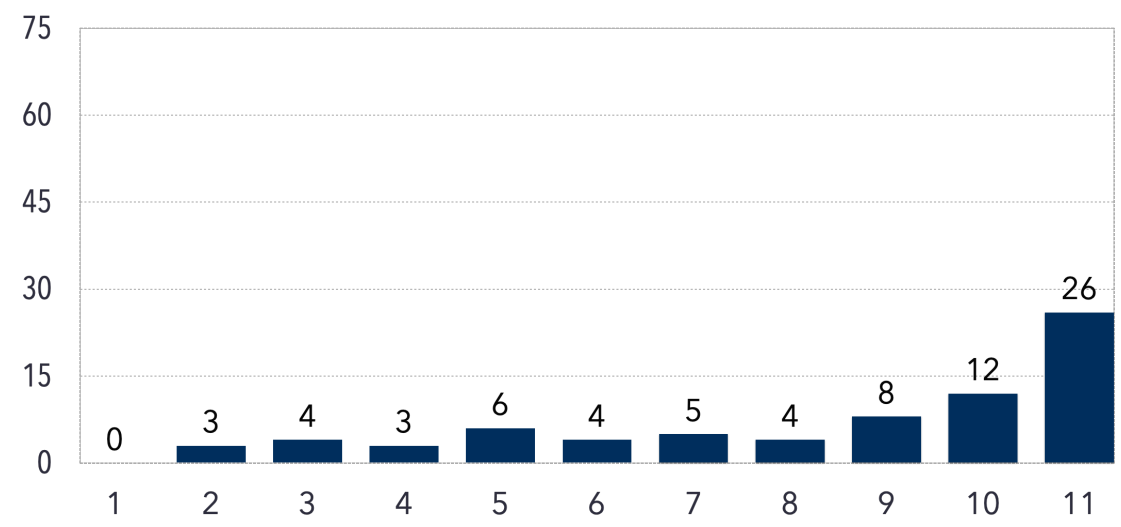 | **9**<br>This statement implies that AIDS does not discriminate and highlights the importance of practicing caution to prevent its transmission. It acknowledges the universal risk of contracting AIDS if proper precautions are not taken. Therefore, it would also be rated positively, around 9, for promoting awareness and emphasizing personal responsibility. |

**Category 11**

**I treat everyone the same, regardless of whether or not they have AIDS**

| **Judges** | | | **ChatGPT** |
|---|---|---|---|
| Median | Interquartile Range | Bar Chart | Category and explanation |
| 11 | 1 | 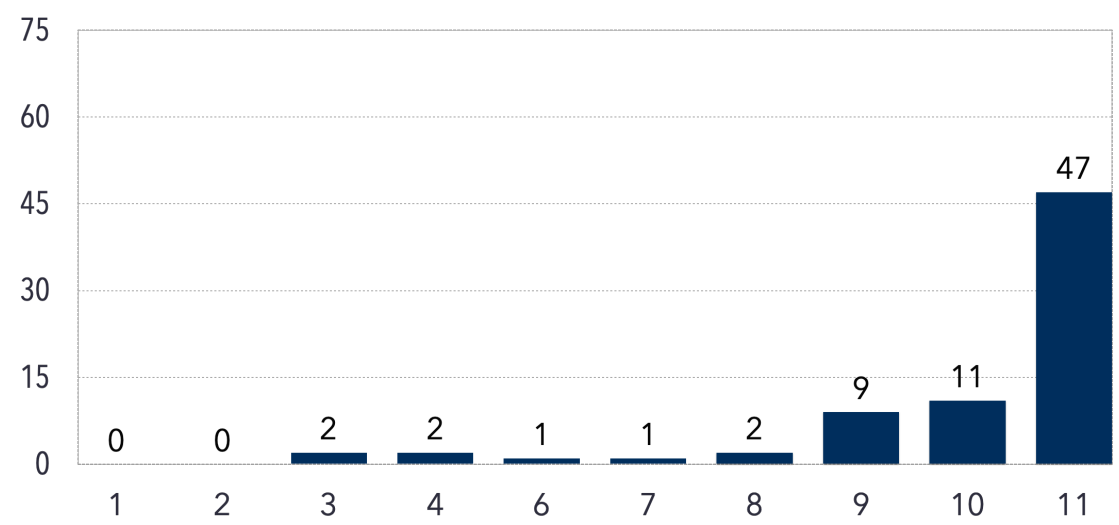 | **9**<br>This statement would rank closer to the positive end of the scale, perhaps 9. It indicates an attitude of non-discrimination towards people with AIDS, emphasizing equality and fair treatment regardless of their health status. |

**AIDS doesn't have a preference, anyone can get it**

| **Judges** | | | **ChatGPT** |
|---|---|---|---|
| Median | Interquartile Range | Bar Chart | Category and explanation |
| 11 | 2 | 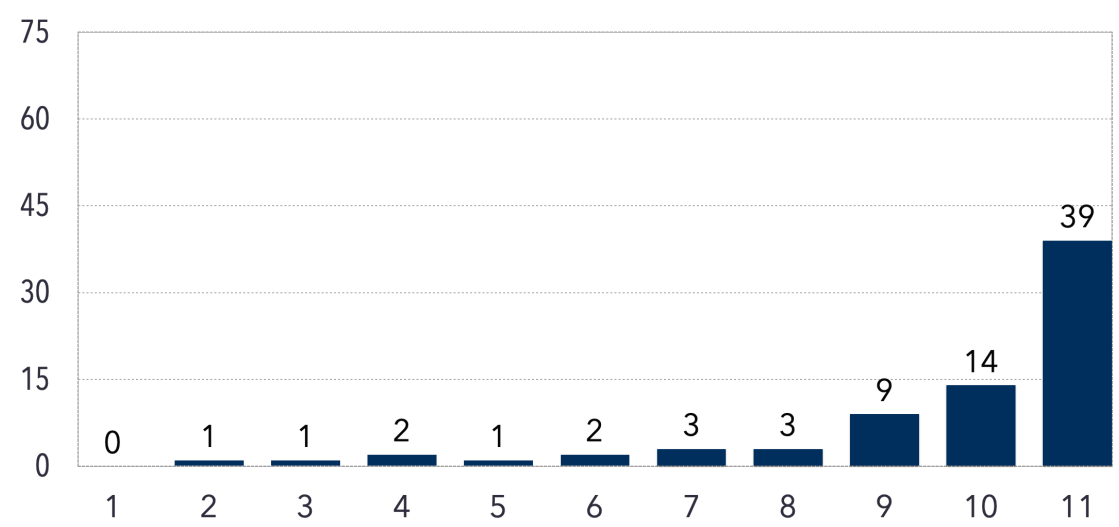 | **9**<br>This statement would likely be rated around 9. It reflects an accurate understanding of HIV/AIDS transmission, emphasizing that the virus does not discriminate based on factors like gender, race, or sexual orientation. |

## 4. Conclusions

In conclusion, our study represents a pioneering effort in methodological innovation for the development of attitude measurement scales, taking as an example an attitude construction towards individuals living with AIDS. By blending traditional psychometric techniques with state-of-the-art Artificial Intelligence tools, we sought to address longstanding challenges in scale construction, such as subjectivity and bias. Through the unique integration of a Large Language Model (LLM) alongside a diverse group of seventy-five human judges, we aimed to achieve a more objective and robust assessment process.

The results of our study reveal promising findings, indicating a significant degree of alignment between the AI and human judges in many instances. This convergence suggests the potential of our integrated approach to not only streamline scale development but also enhance its reliability and validity. By leveraging the computational power and analytical capabilities of modern AI technologies, we have successfully navigated the complexities of attitude measurement, offering new insights and methodologies for future research endeavours.

Furthermore, our study underscores the broader implications of integrating AI into traditional psychometric practices, highlighting the transformative impact of technological advancements on social research methodologies. Beyond the specific context of attitude measurement, our approach opens up exciting possibilities for the application of AI in diverse areas of social science inquiry, paving the way for interdisciplinary collaborations and innovative research methodologies.

In summary, our study represents a significant step forward in the evolution of attitude measurement methodologies, demonstrating the potential of integrating AI technologies to enhance the rigor, efficiency, and objectivity of scale development. As we continue to explore the synergies between human judgment and artificial intelligence, we anticipate further advancements in the field of psychometrics and beyond, ultimately contributing to a deeper understanding of human attitudes and behaviours in the complex socio-cultural landscape. Next steps in our research involve the investigation of the development of specialised algorithms within LLMs tailored specifically for attitude measurement tasks. These algorithms could be trained on large datasets of attitude-related content to improve their accuracy and relevance in categorising attitude items.